\documentclass[journal]{IEEEtran}
\ifCLASSINFOpdf
\else
\fi

\usepackage{amsthm}

\newtheorem{definition}{\textbf{Definition}}
\usepackage{subfigure}
\usepackage{amssymb}
\usepackage{mathtools}
\usepackage{cases}
\usepackage{autobreak}
\usepackage{cuted}
\usepackage{mathrsfs}
\usepackage{booktabs}
\usepackage{multirow}

\usepackage{color}
\usepackage{hyperref}
\usepackage{cite}

\usepackage{amsmath}
\usepackage{extarrows}
\usepackage{amsbsy}
\usepackage{bm}

\usepackage{comment}

\usepackage{algorithmic}
\usepackage{algorithm}
\usepackage{physics}

\newcommand{\E}{\mathbb{E}}

\newcommand{\C}{\mathbb{C}}

\newcommand{\Var}{\mathrm{Var}}

\newcommand{\diag}{\mathrm{diag}}

\newcommand{\BQ}{{\bold{Q}}}
\newcommand{\BH}{{\bold{H}}}
\newcommand{\BI}{{\bold{I}}}
\newcommand{\BX}{{\bold{X}}}

\newcommand{\BS}{{\bold{S}}}
\newcommand{\BZ}{{\bold{Z}}}
\newcommand{\BY}{{\bold{Y}}}

\newcommand{\BG}{{\bold{G}}}
\newcommand{\BJ}{{\bold{J}}}
\newcommand{\Be}{{\bold{e}}}
\newcommand{\Bb}{{\bold{b}}}
\newcommand{\Ba}{{\bold{a}}}
\newcommand{\BA}{{\bold{A}}}
\newcommand{\BB}{{\bold{B}}}
\newcommand{\BC}{{\bold{C}}}
\newcommand{\BD}{{\bold{D}}}
\newcommand{\BW}{{\bold{W}}}
\newcommand{\BM}{{\bold{M}}}

\newcommand{\BL}{{\bold{L}}}
\newcommand{\BR}{{\bold{R}}}
\newcommand{\BT}{{\bold{T}}}

\newcommand{\By}{{\bold{y}}}
\newcommand{\Bn}{{\bold{n}}}
\newcommand{\Bx}{{\bold{x}}}
\newcommand{\Br}{{\bold{r}}}

\newcommand{\Bg}{{\bold{g}}}
\newcommand{\Bh}{{\bold{h}}}
\newcommand{\Bz}{{\bold{z}}}
\newcommand{\Bv}{{\bold{v}}}
\newcommand{\Bq}{{\bold{q}}}

\newcommand{\BP}{{\bold{P}}}
\newcommand{\BN}{{\bold{N}}}

\newcommand{\BV}{{\bold{V}}}
\newcommand{\Bc}{{\bold{c}}}
\newcommand{\Bm}{{\bold{m}}}

\newcommand{\BBA}{\boldsymbol{\mathcal{A}}}
\newcommand{\BBB}{\boldsymbol{\mathcal{B}}}
\newcommand{\BBC}{\boldsymbol{\mathcal{C}}}

\newcommand{\BGamma}{\bold{\Gamma}}
\newcommand{\BOmega}{\bold{\Omega}}
\newcommand{\BTheta}{\bold{\Theta}}
\newcommand{\BSigma}{\bold{\Sigma}}
\newcommand{\BUpsilon}{\bold{\Upsilon}}
\newcommand{\BXi}{\bold{\Xi}}

\newcommand{\Balpha}{\boldsymbol{\alpha}}
\newcommand{\Bbeta}{\boldsymbol{\beta}}

\newcommand{\Blambda}{\boldsymbol{\lambda}}
\newcommand{\BLambda}{\boldsymbol{\Lambda}}

\newcommand{\BDelta}{\boldsymbol{\Delta}}
\newcommand{\BPhi}{\boldsymbol{\Phi}}
\newcommand{\Brho}{\boldsymbol{\rho}}
\newcommand{\Btau}{\boldsymbol{\tau}}
\newcommand{\Bphi}{\boldsymbol{\phi}}
\newcommand{\Bvarsigma}{\boldsymbol{\varsigma}}

\newcommand{\BF}{{\bold{F}}}

\newcommand{\BO}{{\mathcal{O}}}

\DeclareMathOperator{\mino}{minimize}

\newcommand{\RNum}[1]{\uppercase\expandafter{\romannumeral #1\relax}}

\newtheorem{remark}{Remark}
\newtheorem{theorem}{Theorem}
\newtheorem{lemma}{Lemma}
\newtheorem{proposition}{Proposition}
\newtheorem{assumption}{Assumption}
\newtheorem{corollary}{Corollary}
\newtheorem{assumptionalt}{Assumption}[assumption]

\newenvironment{assumptionp}[1]{
  \renewcommand\theassumptionalt{#1}
  \assumptionalt
}{\endassumptionalt}

\usepackage{geometry}
\allowdisplaybreaks[4]
\title{Decentralized MIMO Systems with Imperfect CSI using LMMSE Receivers}
\author{\IEEEauthorblockN {Zeyan Zhuang, \textit{Graduate Student Member}, \textit{IEEE}, Xin Zhang, \textit{Member}, \textit{IEEE}, Dongfang Xu, \textit{Member}, \textit{IEEE}, Shenghui Song, \textit{Senior Member}, \textit{IEEE}, and Yonina C. Eldar, \textit{Fellow}, \textit{IEEE}\vspace*{-10mm}}
\thanks{
Zeyan Zhuang, Xin Zhang, and Shenghui Song are with the Department of Electronic and Computer Engineering,
The Hong Kong University of Science and Technology, Hong Kong (e-mail: zzhuangac@connect.ust.hk; eezhangxin@ust.hk; eeshsong@ust.hk).
\par
Dongfang Xu is with the Division of Integrative Systems and Design,
The Hong Kong University of Science and Technology, Hong Kong (e-mail: eedxu@ust.hk).
\par
Yonina C. Eldar is with the Department of Mathematics and Computer Science,
The Weizmann Institute of Science, Rehovot 7610001, Israel (e-mail: yonina.eldar@weizmann.ac.il).
}
}

\begin{document}

\maketitle

\begin{abstract}
Centralized baseband processing (CBP) is required to achieve the full potential of massive multiple-input multiple-output (MIMO) systems. However, due to the large number of antennas, CBP suffers from two major issues: 1) Extensive data interconnection between radio frequency (RF) circuitry and the central processing unit; and 2)  high-dimensional computation. To this end, decentralized baseband processing (DBP) has been proposed, where the antennas at the base station are partitioned into clusters connected to separate RF circuits and equipped with separate computing units. However, the optimal fusion scheme that maximizes  signal-to-interference-and-noise ratio (SINR) and the related performance analysis for DBP with
general spatial correlation and imperfect channel state information (CSI) have not been studied.  
In this paper, we consider a decentralized MIMO system where all 
clusters adopt linear minimum mean-square error (LMMSE) receivers. We first establish an optimal linear fusion scheme that has high computational and data input/output costs.
To reduce the cost, we then propose two suboptimal fusion schemes with reduced complexity. {For all three schemes, we study the SINR performance  
by leveraging random matrix theory and demonstrate conditions under which the suboptimal schemes are optimal.}
Furthermore, we determine the optimal regularization parameter for the LMMSE receiver, identify the best antenna partitioning strategy, and prove that the SINR will decrease as the number of clusters increases.
Numerical simulations validate the accuracy of the theoretical results.
\end{abstract}


\section{Introduction}
Massive multi-input-multi-output (MIMO) technology has been shown very effective in achieving high spectral efficiency, energy efficiency, 
and link reliability \cite{Lu2014MIMO}. However, conventional implementations require centralized baseband processing (CBP) where all baseband data are collected and processed by a
centralized baseband
processing unit (BBU) \cite{Jeon2019, Vander2018DSP}. 
Such centralized processing causes significant challenges when the size of the antenna arrays becomes extremely large. 
On the one hand, transmitting baseband data from a large number of base station (BS) antennas to the BBU results in very high throughput demands. On the other hand, traditional {estimation} or beamforming algorithms require the inversion of high-dimensional matrices, which causes heightened computational complexity. These challenges make CBP very difficult to implement in practice \cite{Li2017, xu2024distributed}.
\par
To address the bandwidth and computation bottlenecks of CBP, 
a more efficient architecture called decentralized baseband processing (DBP) was 
recently proposed \cite{Li2016, Li2017, Rodr2020, Jeon2019, LiKaipeng2019tradeoff, Dong2022TVT, Zhou2024MCMC}.
With DBP, BS antennas are divided into several clusters, each equipped with its own processing unit that performs signal processing tasks in a parallel and decentralized manner. 
Li \textit{et al}. \cite{Li2016, Li2017} developed iterative decentralized algorithms to minimize the mean-squared error (MSE) of the received symbols. 
 A gradient descent based decentralized algorithm was proposed in \cite{Rodr2020} to obtain zero-forcing (ZF) equalization for a daisy chain architecture. 
 In \cite{Jeon2019, Li2017FPGA}, the feedforward DBP architecture was proposed to reduce latency by avoiding the information exchange between clusters required by previous works \cite{Rusek2013}. 
\par
{With the feedforward structure, the estimates from all clusters and some middle-stage information are transferred to and fused at the central unit (CU). As a result, the fusion algorithm at the CU is crucial for successful estimation.} In \cite{Jeon2019}, Jeon \textit{et al}. provided an optimal linear fusion that maximizes the signal-to-interference-and-noise ratio (SINR) in uncorrelated Rayleigh channels with perfect channel state information (CSI). 
Closed-form deterministic approximations of the SINR for maximum ratio combining (MRC), ZF, and linear minimum mean-square error (LMMSE) receivers were also derived.
The design of the fusion coefficients for decentralized MIMO systems resembles that of the collaboration among access points (APs) in cell-free massive MIMO   (CF-mMIMO) systems, where distributed APs are deployed and linked to a central processing unit (CPU) \cite{Ngo2017mMIMO}. 
The difference between DBP and CF-mMIMO lies in the spatial correlation among different clusters, which makes the DBP architecture more general. In \cite{Björnson2020CFmMIMO}, four levels of cooperation with minimum mean-square error (MMSE) receiver were proposed for CF-mMIMO systems, considering imperfect CSI and spatially correlated
fading. However, the fusion scheme and performance analysis for DBP with imperfect CSI and general spatial correlation have not been studied. 
\par
In this paper, we consider feedforward decentralized massive MIMO systems with LMMSE receivers due to their near-optimal performance and ease of implementation \cite{Rusek2013}.
We also consider the case with imperfect CSI, where the channel estimation (CE) errors are assumed to follow Gaussian distributions \cite{Björnson2020CFmMIMO, Hoydis2013MassiveMIMO}, due to the difficulty in obtaining perfect CSI in massive MIMO systems \cite{ZhangjunICSIT, wei2024resource}. Specifically, we first determine the optimal linear fusion that maximizes the SINR, which requires global CSI and has high complexity. 
To reduce the cost, we propose two linear fusion schemes with reduced computation and communication workloads. In particular,
the first approach utilizes a moderate amount of intermediate results computed by the clusters, while the second method can be implemented by a fully decentralized (FD) architecture \cite{Jeon2019}, where the CU  only needs to calculate the weighted sum of the local estimates, without requiring any other intermediate results. 
\par
Unfortunately, the presence of CE errors and the decentralized structure make it very challenging to evaluate the receive SINR. To this end, we adopt large random matrix theory (RMT) to tackle the random vector channels, which has been proven effective in analyzing the fundamental limits of centralized LMMSE receivers \cite{Abla2009CLTSINR, Abla2009Ber}. The SINR analysis for the DBP architecture has been considered in \cite{Jeon2019} utilizing RMT. However, they assumed perfect CSI and uncorrelated Rayleigh channel, while here we consider the more general case of imperfect CSI and general spatial correlation.
Based on the analysis results, we further investigate the impacts of several key system parameters, including the regularization parameter, the antenna partitioning strategy, and the number of clusters.
\par
The main contributions  can be summarized as follows:
\begin{itemize}
    \item [1)] {For decentralized massive MIMO systems, we first derive the optimal linear fusion scheme to maximize the SINR with imperfect CSI and demonstrate that the optimal coefficients for the fusion scheme also minimize the MSE for estimating the transmit signal. We then propose two suboptimal fusion schemes assuming local CSI, which have reduced levels of complexity.} 
    \item [2)] {We derive the deterministic approximations of the SINR for all three fusion schemes using RMT. The analysis results reveal that, without spatial correlation between clusters or CE errors, the first suboptimal fusion scheme achieves the optimal SINR. The second suboptimal fusion scheme, which has lower complexity, can achieve the optimal SINR when there is no  spatial correlation between clusters.}
    \item [3)] Based on the theoretical analysis, we investigate the scenario with heterogeneous CSI inaccuracy among clusters. The results show that when the CSI error is high for certain clusters, the system's performance is asymptotically equal to that without considering those clusters.
    \item [4)] To obtain more physical insights, we derive closed-form expressions for the SINR over independent and identically distributed (i.i.d.) channels. Based on the results, we further determine the optimal regularization parameters and optimal antenna partition strategies that maximize SINR, and evaluate the impact of the number of clusters.
    \item [5)] From the RMT perspective, we derive the deterministic approximations for linear functions of the resolvent where the covariance matrices have generally correlated columns, and demonstrate the convergence rates of the means are $\BO(N^{-\frac{1}{2}})$. These findings can be applied to the analysis of regularized zero-forcing (RZF) precoding in downlink MIMO systems \cite{Xin2023MISOsecrecy}.
\end{itemize}
\par
The rest of the paper is organized as follows. In Section \ref{Sec_System_model}, we introduce the system model and the DBP architecture. {In Section \ref{Sec_DE}, we derive the optimal linear fusion scheme that maximizes the SINR and propose two linear fusion schemes for decentralized implementation.}  In Section \ref{Sec_AsymSINRAna}, we derive deterministic approximations for the SINR of proposed fusion schemes and investigate the results with i.i.d. channels. Numerical results are given in Section \ref{Sec_Num_Results} and Section \ref{Sec_Conclu} concludes the paper.
\par
\textit{Notations:}
Throughout the paper, lowercase and uppercase boldface letters represent vectors and matrices, respectively. We use $\mathbb{C}^N$ and $\mathbb{C}^{N \times M}$ to denote the space of $N$-dimensional complex vectors and the space of $N$-by-$M$ complex matrices, respectively. 
The conjugate transpose and transpose operator are denoted by $(\cdot)^H$ and $(\cdot)^T$, respectively, 
$[\BA]_{i, j}$ and $[\Ba]_i$ represent the $(i, j)$-th entry of matrix $\BA$ and $i$-th element of $\Ba$, respectively. 
For integer vectors $\Btau = [\tau_1, \ldots, \tau_{N_1}]^T$ and $\Bbeta = [\beta_1, \ldots, \beta_{N_2}]^T$, we use notations $\BA[\Btau, \Bbeta]$ and $\BA[\Btau, :]$ to denote the submatrices of $\BA$ with $[\BA[\Btau, \Bbeta]]_{i, j} = [\BA]_{\tau_i, \beta_j}$ and $[\BA[\Btau, :]]_{i, j} = [\BA]_{\tau_i, j}$, respectively. 
We write $\norm{\cdot}$ for the spectral norm of a matrix or the Euclidean norm of a vector, $\Tr \BA$ refers to the trace of $\BA$, $\BI_N$ represents the identity matrix of size $N$, and $\mathbf{1}_N$ denotes the vector with all 1 entries of size $N$.
The probability measure and 
the expectation operator are denoted by $\mathbb{P}[\cdot]$
 and $\mathbb{E}[\cdot]$, respectively, $\underline{x} = x - \E x$ denotes the centered form of random variable $x$, and $\xrightarrow[]{a.s.}$ indicates almost sure convergence. 
We use $\frac{\partial(\cdot)}{\partial x }$ to represent the partial derivative.
The notations $[N]$ and $[N]_0$ represent the set $\{1, 2, \ldots, N\}$ and $\{0\} \cup [N]$, respectively. 
The indicator function is represented by $\mathbb{I}_{\{\cdot\}}$, and $\mathcal{O}(\cdot)$ is the standard  Big-O notation. Specifically, we have $f(N) = \BO(g(N))$ if and only if there exists a positive real number $C$ and positive integer $N_0$ such that $\abs{f(N)} \leq C g(N)$ for all $N \geq N_0$.
\vspace{-2mm}
\section{System Model}
\label{Sec_System_model}
\subsection{Signal Model}
\begin{figure}[t]
\centering
\includegraphics[width=3.1in]{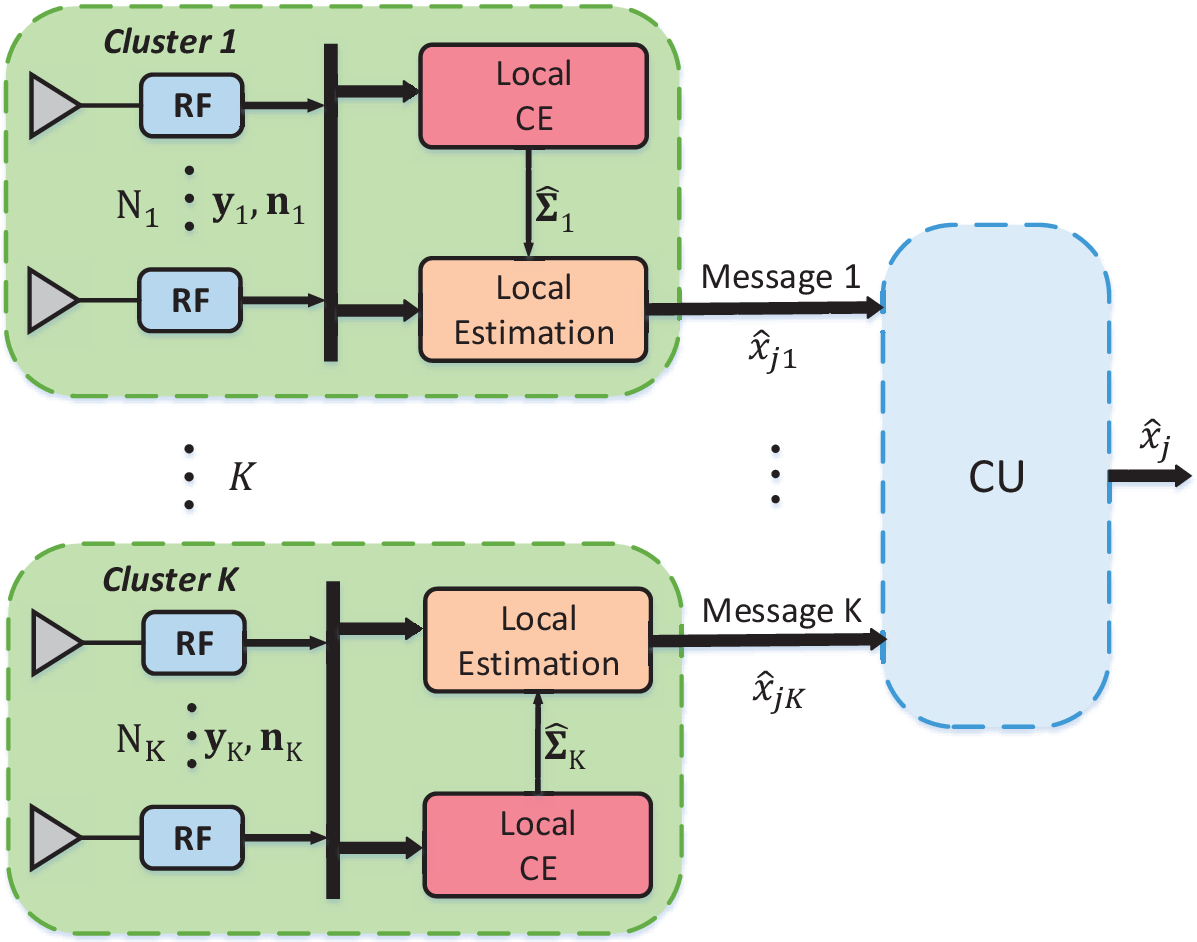}
\caption{The feedforward DBP architecture for massive MIMO uplink.}
\label{Fig_System_model}
\end{figure}
Consider an uplink MIMO system where $M + 1$ single-antenna users transmit signals to one BS with $N$ antennas. 
We denote the transmit signal of all users as $\Bx = [x_0, \ldots, x_{M}]^T \sim \mathcal{CN}(0, \BI_{M+1})$, where $x_j$ represents the signal of user $j$. The channel between user $j$  and the BS is denoted by $\Bh_j \in \mathbb{C}^N$. Thus, the received signal at the BS is given by
\begin{equation}
    \By = \BSigma \Bx + \Bn,
\end{equation}
where $\By \in \mathbb{C}^N$, $\BSigma = [\Bh_0, \ldots, \Bh_{M}] \in \mathbb{C}^{N \times (M+1)}$, and $\Bn \sim \mathcal{CN}(0, \sigma^2\BI_{N})$ denotes the additive Gaussian white noise (AWGN) at the BS. 
\par
\textit{Decentralized Architecture:}
In this paper, we consider the decentralized architecture \cite{Jeon2019, Li2017} as shown in Fig. \ref{Fig_System_model}, where the $N$ antennas of the BS are divided into $K$ clusters, connected to one CU. The $k$-th cluster consists of $N_k$ antennas, and each cluster contains its own RF components and computing unit. For simplicity, we partition the received signal $\By = [\By_1^T, \ldots, \By_K^T ]^T $, the channel matrix $\BSigma = [\BSigma_1^T, \ldots, \BSigma_K^T ]^T $, the channel vector $\Bh_j = [\Bh_{j1}^T, \ldots, \Bh_{jK}^T]^T, j \in [M]_0$, and the noise vector $\Bn = [\Bn_1^T, \ldots, \Bn_K^T ]^T $, such that the received signal at the $k$-th cluster is given by 
\begin{equation}
    \By_k =  \BSigma_k \Bx + \Bn_k, ~ k \in [K],
\end{equation}
with $\By_k \in \mathbb{C}^{N_k}$, $\BSigma_k \in \mathbb{C}^{N_k \times (M + 1)}$, and $\Bn_k \in \mathbb{C}^{N_k}$. 
To estimate the signal from user $j$, the $k$-th cluster independently processes the baseband signal to compute the estimate $\widehat{x}_{jk}$. The local estimate $\widehat{x}_{jk}$ and a moderate amount of intermediate results are then transferred to the CU. These intermediate results are designed based on the estimation scheme and typically include the Gram matrices of the channels\cite{Zhao2023DecLinearprecoding, zhao2023decentralized} or variances of the estimation errors \cite{Jeon2019}. The CU will obtain the final estimate $\widehat{x}_j$ through linear fusion as
\begin{equation}
\label{Eq_LF_alpha_x}
    \widehat{x}_j = \sum_{k=1}^K \alpha_k \widehat{x}_{jk},
\end{equation}
where $\alpha_k$ represents the fusion coefficient that can be chosen as constant or determined based on the information from clusters.  
{To estimate $x_j$, LMMSE estimation is widely used in conventional CBP due to its simplicity in implementation and near-optimal performance \cite{Rusek2013}.}
Therefore, we consider the case where all clusters adopt LMMSE {estimator} and the detailed design for the fusion coefficients $\alpha_k$, $k \in [K]$ will be discussed in Section \ref{Sec_DE}.

\vspace{-1mm}
\subsection{Channel Model}
Due to limited angular spread and insufficient antenna spacing, spatial correlation between BS antennas is inevitable. In this paper, we consider the Rayleigh channel model \cite{Hoydis2013MassiveMIMO, Wagner2012MISO} 
\begin{equation}
\label{Eq_Channel_of_h}
    \Bh_j = \BR_{j}^{\frac{1}{2}} \Bz_j, ~ j \in [M]_0,
\end{equation}
where $\BR_j \in \mathbb{C}^{N \times N}$ is a nonnegative definite matrix that models the spatial correlation of the BS antennas with respect to 
(w.r.t.) user $j$ and $\Bz_j \sim \mathcal{CN}(0, \BI_N)$. Note that this model takes the variance profile model \cite{Abla2009CLTSINR, Hoydis2010Linear} as a special case. With DBP, each cluster can only know its local spatial correlation matrix, which is a principal sub matrix of $\BR_j$. To this end, we denote the index vector
\begin{equation}
    \Bvarsigma_k = \left[\sum_{i=1}^{k-1}N_i + 1,\sum_{i=1}^{k-1}N_i + 2, \ldots,  \sum_{i=1}^{k}N_i\right]^T,~  k \in [K],
\end{equation}
and for a $N$-by-$N$ matrix $\BR$, we define
\begin{equation}
    [\BR]_{[k, l]} = \BR[\Bvarsigma_k, \Bvarsigma_l].
\end{equation}
As a result, the spatial correlation matrix $\BR_j$ can  be expressed in the following block matrix form 
\begin{equation}
\BR_j = \begin{bmatrix}
    [\BR_j]_{[1, 1]} & [\BR_j]_{[1, 2]} & \cdots &[\BR_j]_{[1, K]} \\
    [\BR_j]_{[2, 1]} & [\BR_j]_{[2, 2]} & \cdots & [\BR_j]_{[2, K]} \\
    \vdots & \vdots & \ddots & \vdots  \\
    [\BR_j]_{[K, 1]} & [\BR_j]_{[K, 2]} & \cdots & [\BR_{j}]_{[K, K]}
\end{bmatrix},
\end{equation}
where the off-diagonal block $[\BR_j]_{[k, l]}$ denotes the spatial correlation between the $k$-th and the $l$-th clusters for $k \neq l$, and the main-diagonal block $[\BR_j]_{[k, k]}$ represents the spatial correlation within the $k$-th cluster.

\par
\textit{Decentralized Channel Estimation:}
 For simplicity, we assume $M + 1$ mutually orthogonal pilot signals $\Bphi_j \in \mathbb{C}^{\tau}, j \in [M]_0$ with $\norm{\Bphi_j}^2 = \tau$ ($\tau \geq M+1$). During the training phase, the received signal $\widetilde{\BY}_k \in \mathbb{C}^{N_k \times \tau}$ at the $k$-th cluster  is given by 
\begin{equation}
\label{Eq_cluster_training}
    \widetilde{\BY}_k = \sum_{i=0}^M  \Bh_{ik} \Bphi_i^T + \widetilde{\BN}_k,
\end{equation}
where $\widetilde{\BN}_{k} \in \mathbb{C}^{N_k \times \tau}$ represents the noise matrix with i.i.d. entries, each following $\mathcal{CN}(0, \sigma_{\mathrm{tr}, k}^2)$. To estimate $\Bh_{jk}$, the $k$-th cluster will {first} correlate the training signal with the scaled pilot {$\widetilde{\By}_{jk} = \widetilde{\BY}_k\frac{\Bphi_j^*}{\tau}$}, which yields 
\begin{equation}
\label{Eq_CE_correlated_observation}
\widetilde{\By}_{jk} = \left( \sum_{i=0}^M \Bh_{ik} \Bphi_i^T + \widetilde{\BN}_k \right) \frac{\Bphi_j^*}{\tau} = \Bh_{jk} + \widetilde{\Bn}_{jk},
\end{equation}
where $\widetilde{\Bn}_{jk} \sim \mathcal{CN}(0, \frac{\sigma^2_{\mathrm{tr, k}}}{ \tau })$. Denote the power of the resulting noise as $\widetilde{\sigma}^2_{k} = \frac{\sigma^2_{\mathrm{tr, k}}}{ \tau }$. Thus, the training signal-to-noise ratio (SNR) is $\frac{1}{\widetilde{\sigma}^2_{k}}$, 
which is assumed to be known at each cluster. Under such circumstances, the MMSE estimate $\widehat{\Bh}_{jk}$ of $\Bh_{jk}$ is given by \cite[Theorem 3.1]{bjornson2017massive}
\begin{equation}
    \label{Eq_MMSE_CE}
    \widehat{\Bh}_{jk} = [\BR_{j}]_{[k, k]} \left([\BR_{j}]_{[k, k]} +  \widetilde{\sigma}^2_{k} \BI_{N_k}\right)^{-1} \widetilde{\By}_{jk}.
\end{equation}
Stacking the channel estimate of all clusters as $\widehat{\Bh}_j = [\widehat{\Bh}_{j1}^T, \ldots, \widehat{\Bh}_{jK}^T]^T$, the distribution of the channel vector can be expressed as  
\begin{align}
\label{channel_disttribution}
    \widehat{\Bh}_{j} &\sim \mathcal{CN}(\mathbf{0}, {\BPhi}_j),  \\
    \Bh_j|\widehat{\Bh}_{j} &\sim \mathcal{CN}(\widetilde{\Bh}_j, \BW_j), \notag 
\end{align}
{where the distribution of $\widehat{\Bh}_j$ is derived from the zero mean channel \eqref{Eq_Channel_of_h} and zero mean training noise \eqref{Eq_cluster_training}, using the properties of the linear transformations of Gaussian distributions \cite[Section III]{ComplexGaussian1996}.} {Here, $\BW_j$ represents the covariance matrix of the CE error} and the posterior mean is given by $\widetilde{\Bh}_j = \BV_j \widehat{\Bh}_j$, by the properties of conditional Gaussian distributions \cite[Section IV]{ComplexGaussian1996}.  Matrices $\BPhi_j$, $\BV_j$, and $\BW_j$ are defined as
\begin{align}
\label{Eq_Def_Mat_CE}
    \BPhi_j &=  \BD_{T, j} ( \BD_{\widetilde{\sigma}} + {\BR_j}) \BD_{T, j}, \notag \\
    \BV_j &= \BT_j \BD_{T, j}^{-1}, ~~ \BW_j =  \BT_j\BD_{\widetilde{\sigma}},
\end{align}
where $\BD_{\widetilde{\sigma}} = \diag(\widetilde{\sigma}_{k}^2 \BI_{N_k}; k \in [K])$,  $\BT_j = {\BR_j}( \BD_{\widetilde{\sigma}} + {\BR_j})^{-1}$, $ \BD_{R, j} = \diag([\BR_j]_{[k, k]}; k \in [K])$, and $ \BD_{T, j} =  \BD_{R, j}(\BD_{\widetilde{\sigma}} +  \BD_{R, j})^{-1}$. When the spatial correlation between antennas of different clusters is negligible, i.e., the off-diagonal block $[\BR_j]_{[k, l]} = \mathbf{0}_{N_k \times N_l}$ for $k \neq l$, it can be verified that $\BR_{j} =  \BD_{R, j}$, $\BV_j$ = $\BI_N$, and $\widehat{\Bh}_j$ is the same as that estimated by the centralized MMSE estimator. Therefore, $\BV_j$ indicates the effect of decentralized CE.
 For ease of notation, we define the channel matrices $\widehat{\BSigma} = [\widehat{\Bh}_0, \ldots, \widehat{\Bh}_M]$, $\widehat{\BSigma}_k = [\widehat{\Bh}_{0k}, \ldots, \widehat{\Bh}_{Mk}]$, $\widetilde{\BSigma} = [\widetilde{\Bh}_0, \ldots, \widetilde{\Bh}_M]$, and the sum of the covariance $\BW = \sum_{j \in [M]_0} \BW_j$. 
{\subsection{SINR}
Without loss of generality, we consider the estiomation of $x_0$ using linear reveiver at each cluster. Specifically, the $k$-th
cluster will estimate $\widehat{x}_{0k} = \Br_k^H \By_k$, 
where $
\Br_k = \Br_k\big(\widehat{\BSigma}_k, \{[\BR_j]_{[k, k]}\}_{j=0}^M, \widetilde{\sigma}_k^2\big)
$
depends on the local CSI of the $k$-th cluster. With the linear fusion in 
\eqref{Eq_LF_alpha_x}, the achievable rate of user $0$  can be bounded by a standard lower bound $R_0=  \log\left(1 + \gamma_0 \right)$ based on the worst-case uncorrelated additive noise \cite{2003Howmuchtraining}. Here, $\gamma_0$ is the associated SINR given by 
\begin{equation}
\label{SINR_Exp}
    \gamma_0 = \frac{\left|\E_{\Bh_0|\widehat{\Bh}_0}   \left[ \Br_{\alpha}^{H} {\Bh}_0 \right] \right|^2}{\E_{\BSigma|\widehat{\BSigma}, \Bx, \Bn} \left|\Br_{\alpha}^H  \left[ \BSigma \Bx - \widetilde{\Bh}_0x_0  + \Bn \right] \right|^2 }, 
\end{equation}
with $\Br_{\alpha} = [\alpha_1^*\cdot (\Br_1)^T, \ldots, \alpha_K^* \cdot (\Br_K)^T]^T \in \mathbb{C}^{N}$. 
In this paper, we will determine the optimal fusion coefficients $\alpha_1, \ldots, \alpha_K$ that can maximize the SINR $\gamma_0$ in \eqref{SINR_Exp} \cite{Jeon2019} and 
investigate the asymptotic characterization of $\gamma_0$. 
}
 \section{Decentralized LMMSE Receiver and Linear Fusion}
\label{Sec_DE}
In this section, we first introduce the decentralized LMMSE receiver, and derive the optimal linear fusion scheme which requires global CSI and has high complexity. Then, we propose two linear fusion schemes, which only require local CSI and have reduced complexity. 

\subsection{Decentralized LMMSE Receiver}
\label{Sec_Local_LMMSE_Receiver}
{With perfect CSI, the local MSE of the estimate $\widehat{x}_{0k} = \Br_k^H\By_k$ is given by
\begin{equation}
\mathrm{MSE}_{k} = \E_{\Bx, \Bn}|\widehat{x}_{0k} - x_0|^2,
\end{equation}
where the expectation is w.r.t. the transmit signal and the AWGN. 
However, to obtain the linear receiver that minimizes the local MSE with imperfect CSI, 
the expectation should be performed over  $\BSigma_k | \widehat{\BSigma}_k$ 
\footnote{According to the properties of conditional Gaussian distributions 
\cite[Section IV]{ComplexGaussian1996}, the distribution of the channel vector is given by
$\Bh_{jk} | \widehat{\Bh}_{jk} \sim \mathcal{CN}(\widehat{\Bh}_{jk}, \widetilde{\sigma}_k^2 [\BR_j]_{[k, k]}([\BR_j]_{[k, k]} + \widetilde{\sigma}_k^2\BI_{N_k})^{-1})$.} \cite[Eq. (6)]{Abra2019SINRICSI}, 
which yields \begin{equation}
\label{Eq_Local_LMMSE}
\mathrm{MSE}_{k}|_{\widehat{\BSigma}_k} = \E_{\BSigma_k|\widehat{\BSigma}_k, \Bx, \Bn} \abs{\widehat{x}_{0k} - x_0}^2. 
\end{equation}
Hence, the optimal $\Br_k$ that minimizes $\mathrm{MSE}_{k}|_{\widehat{\BSigma}_k}$ is given by \cite[Proposition 1]{Abra2019SINRICSI}
}
\begin{equation}
    \Br_k^{\mathrm{mmse}} = \left( \widehat{\BSigma}_k \widehat{\BSigma}_k^H + N_k\BZ_k + N_k \rho_k \BI_{N_k}\right)^{-1} \widehat{\Bh}_{0k},
\end{equation}
where  $\rho_k = \frac{\sigma^2}{N_k}$ and
\begin{equation}
\label{Eq_MMSE_parameters_Z}
\BZ_{k} = \frac{\widetilde{\sigma}^2_{k}}{N_k} \sum_{j=0}^M[\BR_j]_{[k, k]}\left([\BR_j]_{[k, k]} + \widetilde{\sigma}^2_{k}\BI_{N_k}\right)^{-1}. 
\end{equation}
Note that the above $\rho_k$ and $\BZ_k$ minimize the local MSE, but may not be optimal after fusion. 
As a result, we treat $\rho_k$ and $\BZ_k$ as design parameters that can be optimized and refer to $\rho_k$ as the regularization parameter in the subsequent.  
\subsection{Linear Fusion with Optimal Coefficients}
The following proposition provides the optimal linear fusion scheme that maximizes the SINR 
and demonstrates that it also minimizes the MSE for estimating the transmit signal.
\begin{proposition}
\label{Prop_opt_alpha}
The optimal coefficients $\Balpha = [\alpha_1, \ldots, \alpha_K]$ for maximizing \eqref{SINR_Exp} 
are given by
\begin{equation}
\label{Eq_opt_alpha}
    \Balpha^{\mathrm{opt}} = c  {\widetilde{\Bh}}_0^H \BD_{r} \left( \BD_{r}^H\left(\widetilde{\BSigma} \widetilde{\BSigma}^H + \BW + \sigma^2 \BI_N\right) \BD_{r}\right)^{-1}, 
\end{equation}
where $c \in \mathbb{C}\backslash \{0\}$ and $\BD_{r} = \diag(\Br_k^{\mathrm{mmse}}; k\in[K])$. When $c= 1$, $\Balpha^{\mathrm{opt}}$ is also the optimal solution for the following MMSE problem 
\vspace*{-2mm}
\begin{equation}
\label{Eq_opt_alpha_MMSE}
    \underset{\Balpha}{\mino} ~~ \mathrm{MSE}|_{\widehat{\BSigma}} = \E_{\BSigma|\widehat{\BSigma}, \Bx, \Bn} \abs{ \sum_{k=1}^K\alpha_k\widehat{x}_{0k} - x_0}^2.
\end{equation}
\end{proposition}
\textit{Proof:} The proof of Proposition \ref{Prop_opt_alpha} is given in Appendix \ref{APP_Prop_opt_alpha}. \qed
\par
\begin{remark}
    The optimality of  $\Balpha^{\mathrm{opt}}$ in Proposition \ref{Prop_opt_alpha} holds for any linear receivers, e.g., MRC, and LMMSE is a special case.
\end{remark}
\begin{remark}
    It can be verified that when $\Balpha = \Balpha^{\mathrm{opt}}$ and $c=1$, we have
    $
        \mathrm{MSE}|_{\widehat{\BSigma}} = (1 + \gamma_0)^{-1} = \frac{\partial \log(1 + \gamma_0)}{\partial \gamma_0}
    $, which reveals the connection
between mutual information and MSE. This relation can be generalized to the case with non-Gaussian signals \cite{Guo2005MMSEGaussian}.
\end{remark}
\begin{remark}
    Proposition \ref{Prop_opt_alpha} shows that the fusion coefficients that maximize the SINR are actually a scaled version of the parameters that minimize MSE. 
\end{remark}
 We will refer to the fusion scheme in Proposition \ref{Prop_opt_alpha} as linear fusion with optimal coefficients (LFOC) in the following. Note that, due to the presence of spatial correlation between different clusters, one cluster cannot determine the off-diagonal elements of $\BR_j -  \BD_{R, j}$ and the channel $\widetilde{\BSigma}$ in the feedforward architecture. 
As a result, to obtain $\Balpha^{\mathrm{opt}}$, the CU needs to know the global CSI, which leads to high input/output (I/O) and computational cost. 
To reduce the complexity, we proposed two suboptimal linear fusion schemes in the following.
\subsection{Linear Fusion with Suboptimal Coefficients}
\label{Sec_Level_2}
{
Based on \eqref{Eq_opt_alpha}, we can design the suboptimal fusion parameter as
\begin{equation}
\begin{split}
\label{Eq_alpha_nopt}
    &\Balpha^{\mathrm{s}\text{-}\mathrm{opt}} =    {\widehat{\Bh}}_0^H\BD_{r}\left( \BD_{r}^H(\widehat{\BSigma} \widehat{\BSigma}^H + \BD_{W} + \sigma^2 \BI_N) \BD_{r}\right)^{-1}, 
\end{split}
\end{equation}
where $\BD_{W} =  \sum_{j \in [M]_0}  \BD_{T, j} \BD_{\widetilde{\sigma}}$. 
}
In particular, the suboptimal solution first replaces $\widetilde{\BSigma}$ in \eqref{Eq_opt_alpha} with the estimated channel $\widehat{\BSigma}$, and then adopts a block diagonal version of $\BW$. These terms are all computed based on the local CSI and spatial correlations. In fact, we can write
\begin{align}
    [\widehat{\BM}]_{k, l} &= (\Br_{k}^{\mathrm{mmse}})^H \widehat{\BSigma}_k\widehat{\BSigma}_l^H \Br_l^{\mathrm{mmse}} \notag\\
    &+ \mathbb{I}_{\{k=l\}}(\Br_k^{\mathrm{mmse}})^H[\BD_{W} + \sigma^2 \BI_{N}]_{[k, k]}\Br_k^{\mathrm{mmse}}, \label{Eq_hat_M}\\
    [\widehat{\Bm}]_k &= (\Br_k^{\mathrm{mmse}})^H\widehat{\Bh}_{0k}, \label{Eq_hat_m}
\end{align}
where $\widehat{\BM} = \BD_{r}^H(\widehat{\BSigma} \widehat{\BSigma}^H + \BD_{W} + \sigma^2 \BI_N) \BD_{r}$ and $\widehat{\Bm} =  \BD_{r}^H{\widehat{\Bh}}_0$. Therefore, to obtain $\Balpha^{\mathrm{s}\text{-}\mathrm{opt}}$, 
the $k$-th cluster needs to compute and transfer the intermediate parameter set
\begin{equation}
\begin{split}
\mathcal{P}_k &= \big\{(\Br_k^{\mathrm{mmse}})^H\widehat{\Bh}_{0k}, ~ (\Br^{\mathrm{mmse}}_k)^H \widehat{\BSigma}_k, \\
&\hspace{8mm}(\Br^{\mathrm{mmse}}_k)^H[\BD_{W} + \sigma^2 \BI_{N}]_{[k, k]}\Br^{\mathrm{mmse}}_k \big\} 
\end{split}
\end{equation}
to the CU. Then the CU will use $\mathcal{P}_1, \ldots \mathcal{P}_K$ to determine $\widehat{\BM}$ and $\widehat{\Bm}$ by  \eqref{Eq_hat_M} and \eqref{Eq_hat_m}, respectively, and compute
 $\Balpha^{\mathrm{s}\text{-}\mathrm{opt}} = \widehat{\Bm}^H\widehat{\BM}^{-1}$. Finally, based on \eqref{Eq_LF_alpha_x}, $\widehat{x}_{0}$ could be obtained. We will refer to this scheme as linear fusion with suboptimal coefficients (LFSC) in the following. Note that the computation and I/O complexity of LFSC is manageable since each parameter in $\mathcal{P}_k$ is a vector or a scalar, and the detailed computational complexity is given in Table \ref{Table_Comlex}.  
It can be observed that the overall complexity of LFSC is significantly reduced compared to CBP and LFOC.
 {
\begin{table*}[!htbp]
    \centering
    \caption{Comparison for the Complexity of Different Estimation Schemes}
    \label{Table_Comlex}
\begin{tabular}{|c|c|c|c| }
\toprule
 Scheme & I/O Throughput of Cluster $k$ &  Computational Complexity of Cluster $k$ & Computational Complexity of the CU\\
\midrule
 Centralized LMMSE & - & - & $C_{\mathrm{L}}(N, M)$ \\ 
 Local LMMSE \& LFOC & $N_kM + 2N_k + 1$ & $C_{\mathrm{L}}(N_k, M)$ & $ N^2M + 2N^2 + NM + 2N + K^2M  + C_\mathrm{F}(K)$ \\
Local LMMSE \& LFSC & $ M + 4$ & $C_{\mathrm{L}}(N_k, M) + N_kM + 2N_k^2 + 2N_k$ & $ K^2M +  C_\mathrm{F}(K)$ \\
  Local LMMSE \& LFCC & $1$ & $C_{\mathrm{L}}(N_k, M)$ & $K$ \\
\bottomrule
\end{tabular}
\vspace{-4mm}
\end{table*} }
\vspace{-4mm}
\subsection{Linear Fusion with Constant Coefficients}
Although LFSC has reduced complexity, it still requires to compute and transmit additional information to the CU. When the number of clusters is very large, the computational complexity, memory requirements, and I/O cost may be very high. To further reduce the complexity, we provide a simple linear fusion scheme with constant coefficients $\alpha_k$. For example, we can take 
\begin{equation}
\label{Eq_LFCC1}
    \alpha_k = \frac{1}{K}, ~~ k \in [K],
\end{equation}
or more heuristically
\begin{equation}
\label{Eq_LFCC2}
    \alpha_k = \frac{N_k}{N}, ~~ k \in [K].
\end{equation}
We will refer to this scheme as linear fusion with constant coefficients (LFCC) in the following. The complexity of LFCC is also given in Table \ref{Table_Comlex} which is much lower than that of LFSC. Note that LFCC can be implemented by the FD feedforward architecture \cite{Jeon2019}, where each cluster only needs to take the weighted sum of local estimates and does not require any additional intermediate results or processing.
\subsection{Complexity Analysis}
The complexities for the proposed schemes and centralized LMMSE, including I/O and computations are summarized in Table \ref{Table_Comlex}. We assume that under LFOC, the CU has access to global statistical CSI. 
\subsubsection{Computational Complexity} The computational complexity primarily arises from multiplication operations. For decentralized estimation, the $k$-th cluster calculates the channel covariance $\widehat{\BSigma}_k{\widehat{\BSigma}}_k^H$, which requires $N^2_k (M+1)$ multiplications. The matrix inversion and the equalization vector $\Br^{\mathrm{mmse}}_k$ require $ N^3_k$ and $N_k^2$ multiplications, respectively. Thus, the total computation complexity for the LMMSE estimate $\widehat{x}_{0k}$ is
\begin{equation}
    C_{\mathrm{L}}(N_k, M) = N_k^2M + N_k^3 + 2 N_k^2 + N_k.
\end{equation}
For LFOC and LFCC, the $k$-th cluster only needs to compute the local estimate which requires $C_{\mathrm{L}}(N_k, M)$ multiplications. The computation of the parameter set $\mathcal{P}_k$ introduces extra $N_kM + 2N_k^2 + 2N_k$ multiplications for LFSC.
With LFOC, the CU computes the channel's posterior mean $\widetilde{\BSigma}$ with $ N^2 (M+1)$ multiplications and the fusion coefficients in \eqref{Eq_opt_alpha} with $N^2 + NM + 2N + K^2M + K^3 + 2K^2$ multiplications. The fusion introduces $K$ multiplications, so the final computational complexity of the CU is given by $ N^2M + 2N^2 + NM + 2N + K^2M + C_\mathrm{F}(K)$ where $C_{\mathrm{F}}(K) =  K^3 + 2K^2 + K$. For LFSC, the CU needs $K^2M + K^3 + 2K^2$  multiplications to obtain $\Balpha^{\mathrm{s}\text{-}\mathrm{opt}}$ based on  $\{\mathcal{P}_k\}_{k=1}^K$. Therefore, the overall complexity of obtaining the final estimate for the CU is $K^2M + C_\mathrm{F}(K)$. For LFCC, $K$ multiplications are used to fuse the local estimates.
\subsubsection{I/O Throughput} For LFOC, the $k$-th cluster needs to transmit the $N_k$-by-$(M+1)$ matrix $\widehat{\BSigma}_k$, the length-$N_k$ vector  $\Br_k^{\mathrm{mmse}}$, and the local estimate $\widehat{x}_{0k}$. LFSC requires $M + 3$ complex scalars in $\mathcal{P}_k$, while LFCC only transmits the local estimate.
\par
We note that in practice, $K$  is much smaller than $N$  and $M$, such that the complexity $C_{\mathrm{F}}(K)$ can be neglected. It can be observed that LFOC has high complexity at the CU. Furthermore, the CU requires global statistical CSI $\{\BR_j\}_{j=0}^M$ for LFOC, resulting in a significant memory overhead. In contrast, LFSC reduces computational complexity at the CU. Overall, the complexity of LFCC is much lower than that of both LFSC and LFOC.
\section{Asymptotic SINR Analysis}
\label{Sec_AsymSINRAna}
The expression for the SINR $\gamma_0$ in \eqref{EQ_gamma_f} contains the correlated terms between the channels $\widehat{\BSigma}$ and $\widetilde{\BSigma}$, making it very difficult to analyze. In this section, we will investigate the asymptotic behaviors of $\gamma_0$.
Specifically, we will show that as the number of antennas and users tend to infinity at the same pace, $\gamma_0$ will converge almost surely. Then, we will derive the deterministic approximations for the SINR of the fusion schemes proposed in Section \ref{Sec_DE}.
To achieve this, we require the following large-scale system assumptions.
\begin{assumptionp}{A.1} \label{A.1}  For a given number of clusters $K$, we assume $0 < \liminf\limits_{N }  \frac{N_k}{M} \leq \limsup\limits_{N } \frac{N_k}{M} < + \infty$ for each $k \in [K]$.
\end{assumptionp}
\begin{assumptionp}{A.2} \label{A.2} Assume 
\begin{subequations}
\begin{align}
     \limsup\limits_N \sup\limits_{j \in [M]_0} \lambda_{\mathrm{M}}(\BR_j) & < + \infty, \\
     \liminf\limits_N \inf\limits_{j\in [M]_0} \lambda_{\mathrm{m}}\left(\BR_j\right) &> 0,
\end{align}
\end{subequations}
where $\lambda_{\mathrm{M}}(\mathbf{R}_j)$ and $\lambda_{\mathrm{m}}(\mathbf{R}_j)$ denote the largest and smallest eigenvalue of $\BR_j$, respectively. 
\end{assumptionp}
\begin{assumptionp}{A.3} \label{A.3} Assume $\rho_k > 0$ and $\BZ_k$ is Hermitian nonnegative such that
    $\limsup\limits_{N} \sup\limits_{k \in [K]} \norm{\BZ_k} < + \infty$ for each $k \in [K]$.
\end{assumptionp}
\textbf{\ref{A.1}} implies that the number of antennas of the $k$-th cluster $N_k$ and the number of users $M$ approach infinity at the same pace \cite{Jeon2019}. For simplicity, we define $c_k = \frac{N_k}{M}$ and use $N \xrightarrow{c_1, \ldots, c_K} + \infty$ to indicate this asymptotic regime. According to interlacing inequalities \cite{marshall1979inequalities}, \textbf{\ref{A.2}} ensures the eigenvalues of the local spatial correlation matrices $[\BR_j]_{[k, k]}$ are bounded away from $0$, which makes $\BV_j$ legal. 
\textbf{\ref{A.3}} implies that the spectral norm of $\BZ_k$ does not diverge to infinity. If \textbf{\ref{A.1}} and \textbf{\ref{A.2}} hold, \eqref{Eq_MMSE_parameters_Z} satisfies this assumption.
\par
 Let $\gamma^{\mathrm{lfoc}}$, $\gamma^{\mathrm{lfsc}}$, 
 and $\gamma^{\mathrm{lfcc}}$
 denote the SINR with LFOC, LFSC, and LFCC, respectively. The following theorem provides the deterministic approximations for $\gamma^{\mathrm{lfoc}}$, $\gamma^{\mathrm{lfsc}}$, and $\gamma^{\mathrm{lfcc}}$, respectively.
\begin{theorem}
 \label{Thm_Asym_SINR_opt}
When Assumptions \textbf{\ref{A.1}}-\textbf{\ref{A.3}} hold, the deterministic approximations for $\gamma^{\mathrm{lfoc}}$, $\gamma^{\mathrm{lfsc}}$,
 and $\gamma^{\mathrm{lfcc}}$ are given by
 \begin{subequations}
 \label{Eq_Thm_sinr_DE}
     \begin{align}
   {\gamma}^{\mathrm{lfoc}} \xrightarrow[N \xrightarrow{c_1, \ldots, c_K} + \infty]{a.s.}\overline{\gamma}^{\mathrm{lfoc}} &= \Bv^H \BDelta^{-1} \Bv, \label{Eq_SINR_LFOC} \\
    {\gamma}^{\mathrm{lfsc}}    \xrightarrow[N \xrightarrow{c_1, \ldots, c_K} + \infty]{a.s.} \overline{\gamma}^{\mathrm{lfsc}} &= \frac{(\Bv^H {\BDelta_I}^{-1} \Bv)^2}{\Bv^H {\BDelta_I}^{-1} \BDelta {\BDelta_I}^{-1} \Bv}, \label{Eq_SINR_LFSC} \\
    {\gamma}^{\mathrm{lfcc}} \xrightarrow[N \xrightarrow{c_1, \ldots, c_K} + \infty]{a.s.}  \overline{\gamma}^{\mathrm{lfcc}} &= \frac{|\Balpha \BJ \Bv|^2}{\Balpha \BJ \BDelta \BJ \Balpha^H}, \label{Eq_SINR_LFCC}
     \end{align}
 \end{subequations}
with
 \begin{subequations}
\begin{align}
    [\Bv]_k &= \frac{\overline{\digamma}_k( [{\BPhi}_0]_{[k, k]})}{N_k} , \\
    \BJ &= \left[\BI_K + \diag(\Bv)\right]^{-1}, \\
    [\BDelta]_{k, l} &=  \frac{\overline{\Upsilon}_{kl}( [\BPhi_{0}]_{[l, k]}, [\BW + \sigma^2 \BI_N]_{[k, l]})}{N_kN_l}    \notag \\
    &+ \frac{\overline{\Pi}_{B, kl}( [\BPhi_{0}]_{[l, k]})}{ \sqrt{N_kN_l}},\\
    [{\BDelta}_I]_{k, l} &=   \frac{\overline{\Upsilon}_{kl}([\BPhi_{0}]_{[l, k]}, [\BD_{W} + \sigma^2 \BI_{N}]_{[k, l]})}{N_kN_l} \notag\\
    &+  \frac{\overline{\Pi}_{kl}( [\BPhi_{0}]_{[l, k]})} {\sqrt{N_kN_l}}, 
\end{align}
 \end{subequations}
where  functions $\overline{\digamma}_k$, $\overline{\Upsilon}_{kl}$, $\overline{\Pi}_{B, kl}$, and $\overline{\Pi}_{kl}$ can be obtained from Lemma \ref{Thm_DE} in Appendix \ref{App_Useful_Results} by setting $z_i = -\rho_i$, $\BS_i = \BZ_i$, $\BA_j = \BPhi_j^{\frac{1}{2}}$, and $\BB_j = \BV_j\BPhi_j^{\frac{1}{2}}$ for each $i \in [K]$ and $j \in [M]$. 
{  
Moreover, the following approximation for the mean holds
\begin{equation}
    \E[\gamma^{\mathrm{fs}}]  =  \overline{\gamma}^{\mathrm{fs}} + \mathcal{O}\left(N^{-1}\right),
\end{equation}
where $\mathrm{fs} \in \{ \mathrm{lfoc}, \mathrm{lfsc}, \mathrm{lfcc}\}$.
}
\end{theorem}
\textit{Proof:} The proof of Theorem \ref{Thm_Asym_SINR_opt} is given in Appendix \ref{APP_Thm_Asym_SINR_opt}. \qed
\begin{remark}
When $K = 1$, which means there is only one cluster, $\overline{\gamma}^{\mathrm{lfoc}}$, $\overline{\gamma}^{\mathrm{lfsc}}$, and $\overline{\gamma}^{\mathrm{lfcc}}$ degenerate to the results with centralized LMMSE \cite{Abla2009CLTSINR, Abla2009Ber}.
\end{remark}
{  
\begin{remark}
Under certain conditions, the performance of LFSC approaches that of LFOC. In particular, when $\BDelta_I = \BDelta$, we have $\overline{\gamma}^{\mathrm{lfoc}} = \overline{\gamma}^{\mathrm{lfsc}}$. The above condition can be achieved when $\BV_j = \BI_N$, for each $j \in [M]_0$, which is equivalent to 
\begin{equation}
\label{Eq_cond_lfsc_equ_lfoc}
\BD_{\widetilde{\sigma}} \BD_{R, j} = \BD_{\widetilde{\sigma}} \BR_j.
\end{equation}
Denote the index set of the clusters with perfect CSI as $\mathcal{K}_1 \subset [K]$ such that $\widetilde{\sigma}_{k_1}^2 = 0$, for $ k_1 \in \mathcal{K}_1$. The above condition \eqref{Eq_cond_lfsc_equ_lfoc} can be achieved if $[\BR_j]_{[k, l]} = \mathbf{0}_{N_k \times N_l}$  when the indexes $k$ and $l$ ($k \neq l$) do not belong to the set $\mathcal{K}_1$ simultaneously. Furthermore, when the spatial correlations among all clusters are negligible, i.e., $\BR_j =  \BD_{R, j}$ for each $j \in [M]_0$ or the CSI is perfect for all of the clusters, i.e., $\widetilde{\sigma}^2_k = 0$ for each $k \in [K]$, \eqref{Eq_cond_lfsc_equ_lfoc} is satisfied automatically.
\end{remark}
}
\begin{remark}
\label{Rm_LFCC_Opt}
{ 
     When there is no spatial correlation between clusters, the matrix $\BDelta$ becomes diagonal. Under such circumstances, it can be verified that when $\Balpha = C [\frac{(1 + [\Bv]_1) [\Bv]_1}{[\BDelta]_{1, 1}}, \ldots, \frac{(1 + [\Bv]_K) [\Bv]_K}{[\BDelta]_{K, K}}]$ with $C \neq 0$ being an arbitrary constant in \eqref{Eq_SINR_LFCC}, we have 
     \begin{equation}
     \overline{\gamma}^{\mathrm{lfoc}} = \overline{\gamma}^{\mathrm{lfsc}} = \overline{\gamma}^{\mathrm{lfcc}} = \sum_{k=1}^K \frac{[\Bv]_k^2}{[\BDelta]_{k, k}}, 
     \end{equation}
     which is equal to the sum of the SINRs from all clusters.} This is consistent with
    the conclusion of \cite[Theorem 6]{Jeon2019}, which assumes uncorrelated Rayleigh channels and perfect CSI.
     In addition, $[\Bv]_{k}$ and $[\BDelta]_{k, k}$ are only related to the local spatial correlation, which varies slowly in time. As a result, the fusion coefficients for LFCC can be set as $\alpha_k = \frac{(1 + [\Bv]_k) [\Bv]_k}{[\BDelta]_{k, k}}$
  to achieve the asymptotically optimal performance.
\end{remark}
{  
\begin{remark}
    By applying Taylor expansion, it can be verified that the achievable rate can be calculated as 
    $\mathbb{E}\left[\log(1 + \gamma^{\mathrm{fs}})\right] = \log(1 + \overline{\gamma}^{\mathrm{fs}}) + \mathcal{O} (N^{-1})$ 
    ,where $\mathrm{fs} \in \{ \mathrm{lfoc}, \mathrm{lfsc}, \mathrm{lfcc}\}$. 
    These results can be utilized to evaluate the ergodic rate.
\end{remark}
}
{ 
\begin{remark}
    The result in \eqref{Eq_Thm_sinr_DE} can be significantly simplified and given in a closed form with i.i.d. channels, providing useful insights for system design. More details are available in Section \ref{Sec_IID_Channel}.  It will be shown that there is a trade-off between complexity and performance. It is recommended to adopt LMMSE estimator locally, minimize the number of clusters, and avoid uniform partitioning.
\end{remark}
}
{ Next, we investigate the case with heterogeneous CSI inaccuracy levels among the clusters. Specifically, the following corollary illustrates the behavior of the SINR when certain clusters have severe CSI errors.
\begin{corollary}
\label{Coro_Unbalance_CSI}
     Let $\mathcal{K}_e$ be the index set of clusters with severe CSI errors, i.e.,  $\widetilde{\sigma}^2_k \to +\infty$ for $k \in \mathcal{K}_e$, while $\widetilde{\sigma}^2_l$ is fixed for $l \in \mathcal{K}_g = [K] - \mathcal{K}_e = \left\{ k_1, k_2, \ldots, k_{G} \right\}$, where $G = K - \abs{\mathcal{K}_e}$. Then the deterministic approximation of the SINR satisfies
     \begin{equation}
     \label{Eq_Equi_SINR_CSI_unbalance}
         \overline{\gamma}^{\mathrm{fs}} \rightarrow \overline{\gamma}^{\mathrm{fs}}_{\mathcal{K}_g},
     \end{equation}
     where $\mathrm{fs} \in \{ \mathrm{lfoc}, \mathrm{lfsc}, \mathrm{lfcc}\}$ and $\overline{\gamma}^{\mathrm{fs}}_{\mathcal{K}_g}$ is obtained by replacing the channel $\Bh_j$ by $\Bh_{j}^{\mathcal{K}_g} = [\Bh_{jk_1}^T, \ldots, \Bh_{jk_G}^T]^T$ for each $j \in [M]_0$ and $\Balpha$ by $\Balpha^{\mathcal{K}_g} = [\alpha_{k_1}, \ldots, \alpha_{k_G}]$ in Theorem \ref{Thm_Asym_SINR_opt}.
\end{corollary}
\textit{Proof:} The proof of Corollary \ref{Coro_Unbalance_CSI} is given in Appendix \ref{App_Proof_Coro_Unbalance_CSI}. \qed
\begin{remark}
Corollary \ref{Coro_Unbalance_CSI} indicates that for decentralized estimation, when several clusters have severe CSI errors, the performance will be equal to that of a system where these clusters are ignored.
\end{remark}
}
\subsection{Special Case Study: I.I.D. Channels}
\label{Sec_IID_Channel}
In this section, we will consider the simplified channel model with $\BR_j = \BI_N$ to get some physical insights. The parameter $\BZ_k$ is set as \eqref{Eq_MMSE_parameters_Z}, i.e., $ \BZ_k = \frac{(M+1)\widetilde{\sigma}^2_k}{N_k( \widetilde{\sigma}^2_k + 1)} \BI_{N_k}$. 
\subsubsection{SINR Approximation} With i.i.d. channels, the SINR in Theorem \ref{Thm_Asym_SINR_opt} can be given in a closed-form.
{ 
\begin{corollary}
\label{IID_Case}
    Given assumptions \textbf{\ref{A.1}}-\textbf{\ref{A.3}} and the i.i.d. channels,  $\overline{\gamma}^{\mathrm{lfoc}}$, $\overline{\gamma}^{\mathrm{lfsc}}$ and $\overline{\gamma}^{\mathrm{lfcc}}$ can be given by
    \begin{subequations}
    \begin{align}
        \overline{\gamma}^{\mathrm{lfoc}} &= \overline{\gamma}^{\mathrm{lfsc}} =  \sum_{k=1}^K \frac{\delta_k }{\varpi_k},  \label{gamma_opt}\\
        \overline{\gamma}^{\mathrm{lfcc}}&= \frac{\abs{\sum_{k=1}^K \alpha_k \frac{\delta_k}{1 + \delta_k}}^2}{\sum_{k=1}^K\frac{|\alpha_k|^2 \delta_k}{(1 + \delta_k)^2}\varpi_k},  \label{iid_gamma_const}
    \end{align}
    \end{subequations}
where 
\begin{align}
     \varpi_k = 1 + \left(\frac{\sigma^2}{Mc_k} - \rho_k \right) \frac{(\widetilde{\sigma}^2_k + 1) \delta_k}{1 -  \frac{\delta_k^2}{c_k(1 + \delta_k)^2}}
\end{align}
and $\delta_k$ is given by
\begin{equation}
\label{Eq_iid_delta}
    \delta_k = \frac{1 - \frac{1}{c_k} - A_k + \sqrt{\left(A_k + \frac{1}{c_k} - 1\right)^2+ 4A_k} }{2A_k}
\end{equation}
with $A_k = \rho_k(\widetilde{\sigma}^2_k +1 ) + \frac{M+1}{Mc_k} \widetilde{\sigma}^2_k$.
\end{corollary}
}

\textit{Proof:} The proof can be obtained by setting $\BR_j = \BI_N$ and $\BZ_k = \frac{(M+1) \widetilde{\sigma}^2_k}{N_k(\widetilde{\sigma}^2_k + 1)}$ in Theorem \ref{Thm_Asym_SINR_opt} and is omitted here. \qed

{ 
\begin{remark}
    According to the definition, it can be verified $\delta_k = [\Bv]_k$ and $\varpi_k = \frac{[\BDelta]_{k, k}}{[\Bv]_k}$ in Theorem \ref{Thm_Asym_SINR_opt}.
\end{remark}
}
\par
Denote $\Brho_K = [\rho_1, \ldots, \rho_K]^T$ and $ \Bc_K = [c_1, \ldots, c_K]^T$. We next investigate the impact of parameters $\Brho_K$, $\Bc_K$, and the number of clusters $K$ on the SINR $\overline{\gamma}^{\mathrm{lfoc}} = \overline{\gamma}^{\mathrm{lfoc}}_K(\Brho_K, \Bc_K)$ where we keep $N$ and $M$ fixed.  {  We also assume the CSI inaccuracy level is identical among clusters, i.e., $\widetilde{\sigma}_k^2 = \widetilde{\sigma}^2$ for each $k \in [K]$ in the following. }

\subsubsection{Regularization Parameter} The following corollary gives the optimal parameter $\Brho_K$.
\begin{corollary}
\label{Coro_best_rho}
    Assume the same conditions as Corollary \ref{IID_Case}. For any $\Brho_K > 0$, $\Bc_K \geq 0$, we have 
    \begin{equation}
\label{Eq_best_rhoK}\overline{\gamma}^{\mathrm{lfoc}}_K(\Brho_K, \Bc_K) \leq \overline{\gamma}^{\mathrm{lfoc}}_K\left(\Brho^{\mathrm{opt}}_K, \Bc_K \right),
    \end{equation}
    where $\Brho^{\mathrm{opt}}_K = [\frac{\sigma^2}{Mc_1}, \ldots, \frac{\sigma^2}{Mc_K}]^T$.
\end{corollary}
\textit{Proof:} The proof of Corollary \ref{Coro_best_rho} is given in Appendix \ref{App_Coro_best_rho}. \qed
\par
\begin{remark}
    We can observe that the optimal parameter $\Brho_K^{\mathrm{opt}}$ that maximizes the SINR is identical to that achieving the minimal local MSE as derived in Section \ref{Sec_Local_LMMSE_Receiver}. This can be intuitively obtained from \eqref{gamma_opt} where the SINR $\overline{\gamma}^{\mathrm{lfoc}}$ is the sum of the SINRs of all clusters.
\end{remark}
\subsubsection{Antenna Partitioning Strategy} The following corollary shows the impact of the parameter $\Bc_K$.
\begin{corollary}
\label{Coro_best_c}
    Assume the same conditions as Corollary \ref{IID_Case} and denote $\Brho_K = [\frac{a}{c_1},  \ldots, \frac{a}{c_K}]^T$. For any $\Bc_K \geq 0$ and $  \frac{\sigma^2}{M} \leq a$, we have
    \begin{equation}
    \label{Eq_gamma_c_M}
         \overline{\gamma}^{\mathrm{lfoc}}_K(\Brho_K, \Bc_K) \leq \overline{\gamma}^{\mathrm{lfoc}}_K(\Brho_K, {\Bc}_{K, \mathrm{M}}),
    \end{equation}
    and for any $\Bc_K \geq 0$ and $\frac{\sigma^2}{M}  \leq a \leq \frac{2\sigma^2}{M} + \frac{(M+1) \widetilde{\sigma}^2}{M(\widetilde{\sigma}^2 + 1)}$, we have
    \begin{equation}
    \label{Eq_c_m}
    \overline{\gamma}^{\mathrm{lfoc}}_K\left(\Brho_K, \Bc_{K, \mathrm{m}}\right) \leq \overline{\gamma}^{\mathrm{lfoc}}_K\left(\Brho_K, \Bc_K \right),
    \end{equation}
    where $\Bc_{K, \mathrm{m}} = [\overline{c}, \ldots, \overline{c}]^T$, $\Bc_{K, \mathrm{M}} = [K \overline{c}, 0, \ldots, 0]^T$, and $\overline{c} = \frac{1}{K}\sum_{k=1}^K c_k$.
\end{corollary}
\textit{Proof:} The proof of Corollary \ref{Coro_best_c} is given in Appendix \ref{App_Coro_best_c}. \qed
\begin{remark}
Inequalities $\eqref{Eq_gamma_c_M}$ and \eqref{Eq_c_m} imply that when the regularization parameter $\Brho_K$ is within a certain range, i.e., $\frac{\sigma^2}{M} \leq a$, the system achieves the maximum SINR 
when the receive antennas are concentrated in one cluster, i.e., the centralized scheme. On the other side, when the antennas are equally partitioned, the system achieves the lowest SINR. This finding aligns with the conclusions in \cite[Lemma 7, Lemma 8]{Jeon2019}. Note that \cite{Jeon2019} considers a given parameter $\Brho_K$ with perfect CSI, but Corollary \ref{Coro_best_c} provides the range for $\Brho_K$ with imperfect CSI.
\end{remark}
\begin{remark}
    It can be observed that the optimal regularization parameter $ \Brho^{\mathrm{opt}}_K$ in Corollary \ref{Coro_best_rho} is within the range such that $\eqref{Eq_gamma_c_M}$ and \eqref{Eq_c_m} hold.
\end{remark}
\subsubsection{Number of Clusters} Corollary \ref{Coro_best_c} compares different antenna partitioning strategies with a fixed number of clusters $K$. Another interesting question to investigate is how the number of clusters $K$ affects the SINR. The following corollary demonstrates this impact.
\begin{corollary}
\label{Coro_impact_K}
    Assume the same conditions as Corollary \ref{IID_Case} and denote $\Brho_K = [\frac{a}{c_1}, \ldots, \frac{a}{c_K}]^T$. For any $\frac{\sigma^2}{M} \leq a$, we have
    \begin{equation}
        \label{Eq_Mono_SINR_K}\overline{\gamma}^{\mathrm{lfoc}}_K(\Brho_K, \Bc_{K, \mathrm{m}}) \geq \overline{\gamma}^{\mathrm{lfoc}}_{K+1}(\Brho_{K+1}, \Bc_{K+1, \mathrm{m}}).
    \end{equation}
\end{corollary}
\textit{Proof:} The proof of Corollary \ref{Coro_impact_K} is given in Appendix \ref{App_Coro_impact_K}. \qed
\begin{remark} Since the total number of receive antennas $N$ is fixed, Corollary \ref{Coro_impact_K} indicates that as the number of clusters increases, the SINR will monotonically decrease.    
\end{remark}
\begin{remark}
    It can be verified from the proof of Corollary \ref{Coro_impact_K} that 
    \begin{equation}
    \label{Eq_SINR_K_LowerBd}
\overline{\gamma}^{\mathrm{lfoc}}_K(\Brho_K, \Bc_{K, \mathrm{m}}) \geq \frac{N}{(\sigma^2 + M)(\widetilde{\sigma}^2 + 1) + \widetilde{\sigma}^2}, 
    \end{equation}
    and the equality holds when $K \rightarrow + \infty$, which is not achievable since $K \leq N$.
    From Corollary \ref{Coro_best_c}, we know that the equal antenna partitioning strategy has the worst performance. Therefore, for decentralized LMMSE receivers, the SINR is strictly lower bounded by the right hand side (RHS) of \eqref{Eq_SINR_K_LowerBd}. When the numbers of antennas and users are large, this lower bound is approximately $\frac{N}{M}$.
\end{remark}
\section{Numerical Results}
\label{Sec_Num_Results}
In this section, we will verify the accuracy of the theoretical results in Sections \ref{Sec_AsymSINRAna}. We will also demonstrate the impact of several key system parameters over i.i.d. channels, including the regularization parameter, antenna partitioning strategy, and the number of clusters. For simplicity, we use the notation $\mathcal{N}_K = (N_1, \ldots, N_K)$ to denote the antenna partitioning strategy. The Monte Carlo (MC) simulation results were obtained from 5000 realizations and are illustrated by markers in all figures.
\subsection{Correlated Channels}
We consider a uniform linear array of antennas at the BS. The $(m, n)$-th entry of the spatial correlation matrix at BS is modeled as \cite{Zhangjun2013Large, Xin2023Secrecy}
\begin{equation}
\begin{split}
    &\left[\mathbf{C}(\eta, \delta,  d_s)\right]_{m, n} \\
    &= \int_{-180^{\circ}}^{180^{\circ}} \frac{1 }{\sqrt{2 \pi \delta^2}} e^{2\pi\jmath d_s (m-n) \sin(\frac{\pi \phi}{180}) - \frac{(\phi - \eta)^2}{2 \delta^2}} \mathrm{d} \phi.
\end{split}
\end{equation}
Here, $d_s$ and $\phi$ denote the receive antenna spacing and angular spread, respectively, $\eta$ represents the mean angle, and $\delta$ denotes the root-mean-square angle spread. In the simulation, we set $\BR_j = \BC(\eta_j, \delta_j, 1)$ with $\eta_j = (\frac{j}{180 M})^{\circ}$ and $\delta_j = 10^\circ + (\frac{j}{M10})^{\circ}$.  The parameter $\rho_k$ is set as $\rho_k = \frac{\sigma^2}{N_k}$ and $\BZ_k$ is set according to \eqref{Eq_MMSE_parameters_Z} for each $k \in [K]$.
\par
\begin{figure}[t!]
\centering
\vspace{-2mm}
 \subfigure[  SINR versus signal SNR $\frac{1}{\sigma^2}$ with training SNR $\frac{1}{\widetilde{\sigma}^2_1} = \frac{1}{\widetilde{\sigma}^2_2} = 30$ dB.]{
\includegraphics[width=3.1in]{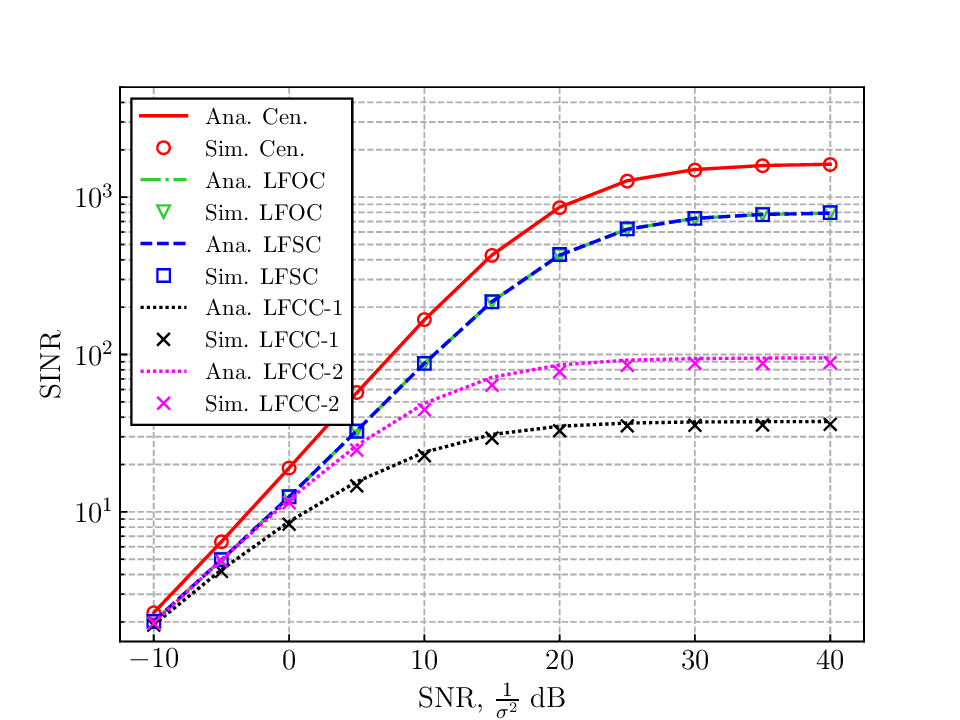}
\label{Fig1(a)}
}

 \subfigure[{  SINR versus training SNR $\frac{1}{\widetilde{\sigma}_1^2} = \frac{1}{\widetilde{\sigma}_2^2} = \frac{1}{\widetilde{\sigma}^2}$ with signal SNR $\frac{1}{{\sigma}^2} = 30$ dB.}]
 {\vspace*{-4mm}
\includegraphics[width=3.1in]{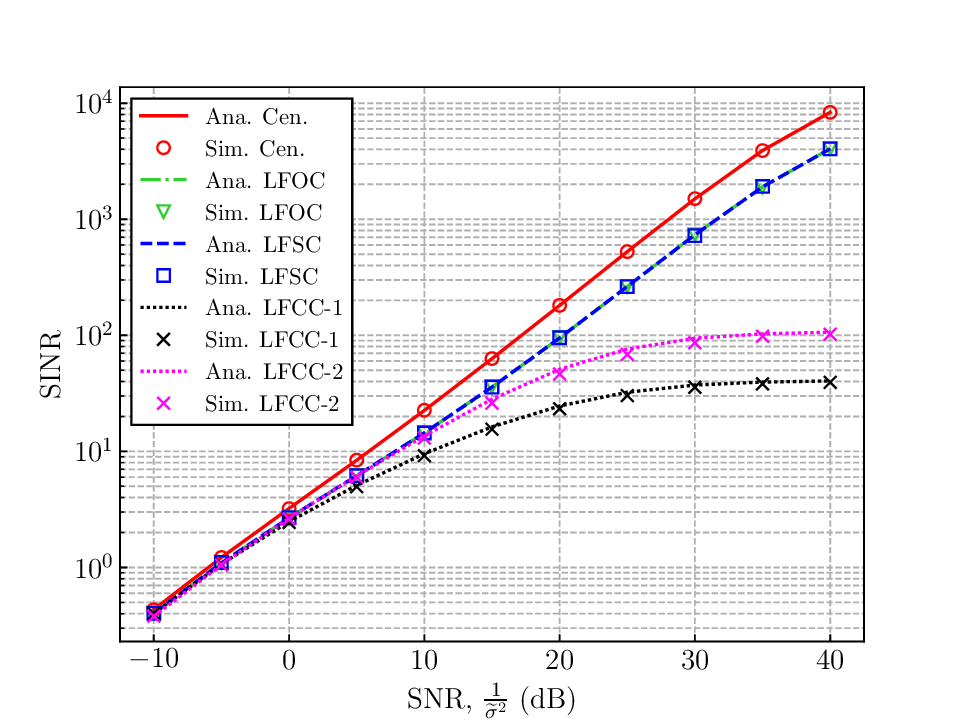}
\label{Fig1(b)}
}
\caption{SINR with different linear fusion schemes.} 
\label{Fig1_SINR}
\end{figure}

{ \textbf{Approximation of SINR: } Fig. \ref{Fig1_SINR} illustrates the SINR with different linear fusion schemes. The system dimensions are set as $N = 32$, $M = 12$, $K = 2$, and $\mathcal{N}_2 = (10, 22)$. The red line and the two dotted lines represent the results of centralized LMMSE and LFCC, respectively. The fusion parameter is set as \eqref{Eq_LFCC1} for LFCC-1 and \eqref{Eq_LFCC2} for LFCC-2, respectively. It is shown that centralized LMMSE performs better than the three decentralized schemes. By comparing the results with MC simulations, we can observe that the approximation results in Theorem \ref{Thm_Asym_SINR_opt} are accurate.  By comparing the results of LFSC and LFOC, it can be observed that LFSC is highly efficient. Furthermore, Fig. \ref{Fig1_SINR} shows that when the antenna partition is unbalanced, using LFCC will lead to performance degradation.}
\par
\begin{figure}[t]
\centering
\vspace{-4mm}
\includegraphics[width=3.1in]{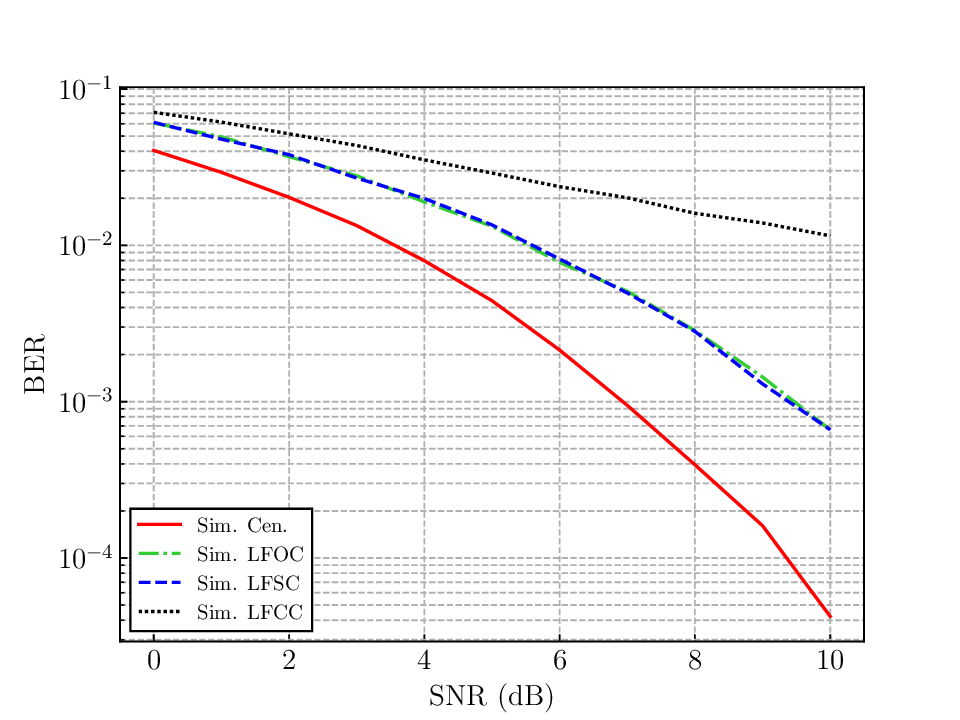}
\caption{  BER versus SNR.} 
\label{Fig_BER}
\end{figure}

{ 
\textbf{BER Performance:} Fig. \ref{Fig_BER} shows the bit error rate (BER) with 16-QAM modulation for different fusion schemes. The system parameters are set as $N = 108$, $M = 52$, $K = 2$, and $\mathcal{N}_2 = (20, 88)$. The fusion parameter for LFCC is set as $\Balpha = [0.5, 0.5]$ and the signal SNR and training SNR of two clusters are set to be equal, with $\frac{1}{\sigma^2} = \frac{1}{\widetilde{\sigma}^2_1} = \frac{1}{\widetilde{\sigma}^2_2}$, represented by the horizontal axis in Fig. \ref{Fig_BER}. It can be observed that the BER for centralized architecture is much lower than that with the decentralized scheme.}
\par
\begin{figure}[t]
\centering
\vspace{-3mm}
\includegraphics[width=3.1in]{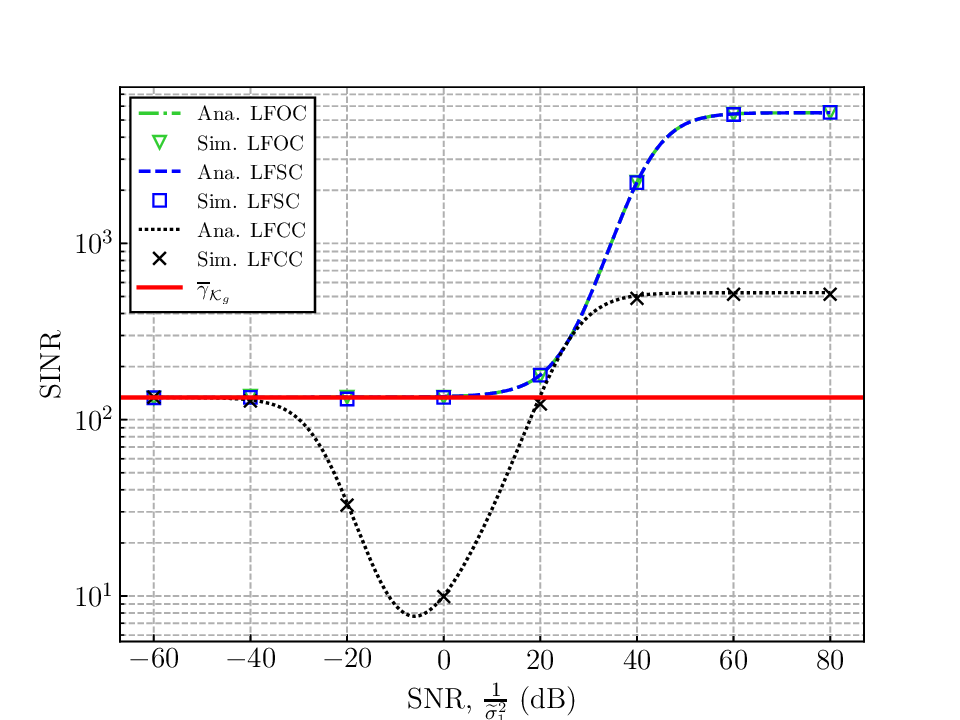}
\caption{  SINR versus training SNR $\frac{1}{\widetilde{\sigma}_1^2}$.} 
\label{Fig_unbalance_SNR}
\end{figure}

{  \textbf{Heterogeneous CSI Inaccuracy:}
 Fig. \ref{Fig_unbalance_SNR} illustrates the performance of SINR with different levels of CSI inaccuracy among clusters. The system parameters are set as $N = 40$, $M = 16$, $K = 2$, $\mathcal{N}_2 = (22, 18)$, and $\frac{1}{\sigma^2} = \frac{1}{\widetilde{\sigma}^2_2} = 30$ dB. The fusion parameter for LFCC is set as $\Balpha = [0.5, 0.5]$. The red line represents the performance of centralized estimation considering only the second cluster. It can be observed that when the training SNR of the first cluster is very low, the performance approaches that without considering the first cluster, which validates Corollary \ref{Coro_Unbalance_CSI}.}
 \par
\begin{figure}[t]
\centering
\vspace{-2mm}
\includegraphics[width=3.2in]{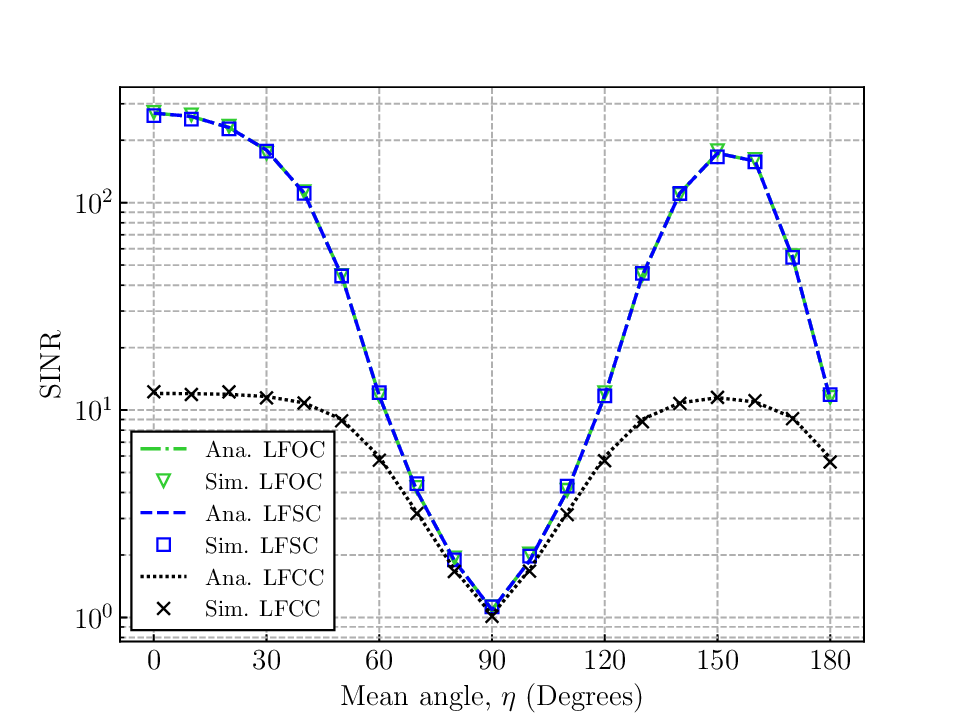}
\caption{  SINR versus mean angle $\eta$. } 
\label{Fig_SINR_Mean_Angle}
\end{figure}

{ 
\textbf{Impact of Spatial Correlations:}  Fig. \ref{Fig_SINR_Mean_Angle} illustrates the impact of the mean angle of the spatial correlation matrices on SINR for different fusion schemes.  The spatial correlations are set to be $\BR_j = \BC(\eta, 20, 1)$. The parameters are set as $N = 40$, $M = 18$, ${\mathcal{N}}_2 = (28, 12)$, and $\frac{1}{\sigma^2} = \frac{1}{\widetilde{\sigma}^2_1} = \frac{1}{\widetilde{\sigma}^2_2} = 30$ dB. For LFCC, $\Balpha$ is set as $[0.5, 0.5]$.  It can be observed that when $\eta = 90^{\circ}$, the SINR is the lowest. This is because when $\eta = 90^\circ$, the spatial correlation is strong, leading to many eigenvalues of $\BR_j$ being close to zero.
}

\par
\begin{figure}[t]
\centering
\vspace*{-2mm}
\includegraphics[width=3.1in]{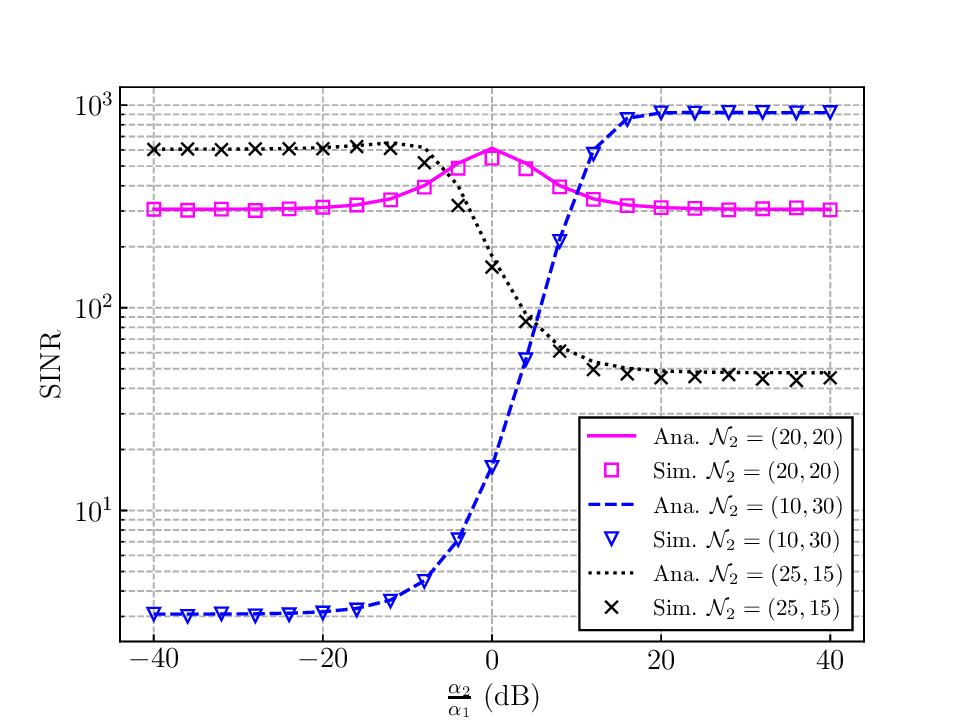}
\vspace*{-2mm}
\caption{SINR versus the ratio of fusion coefficients $\frac{\alpha_2}{\alpha_1}$.} 
\vspace{-1mm}
\label{Fig3}
\end{figure}

\textbf{Impact of Fusion Coefficients: }Fig. \ref{Fig3} shows the impact of the fusion coefficients with different antenna partitioning strategies. The system dimensions are set as $N = 40$, $M = 15$, and $K = 2$. The signal and training SNR are both set as $30$ dB. It can be observed that when antennas are uniformly allocated to different clusters, i.e., $\mathcal{N}_2 = (20, 20)$, SINR is not sensitive to the value of $\Balpha$, but the maximum SINR is lower than that with unbalanced antenna partitioning. This observation for correlated channels is consistent with the conclusion for i.i.d. channels in Corollary \ref{Coro_best_c}. Moreover, as the number of antennas in a cluster increases, the corresponding linear fusion weights should also increase.
\subsection{I.I.D. Channels}
In the experiment, the training and signal SNR are both 30 dB, and the parameter $\BZ_k$ is set according to $\eqref{Eq_MMSE_parameters_Z}$ for each $k \in [K]$.
\par
\begin{figure}[t]
\centering
\includegraphics[width=3.1in]{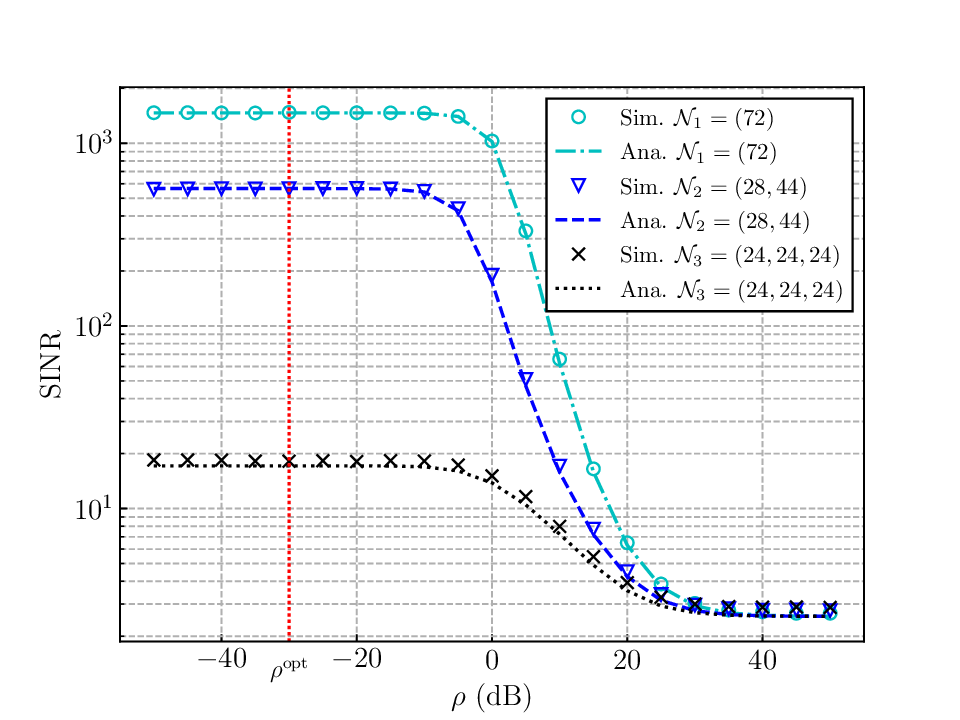}
\caption{SINR versus the regularization parameter $\rho$.} 
\label{Figiid_1}
\end{figure}

\textbf{Optimal Regularization Parameter:} 
Fig. \ref{Figiid_1} depicts the impact of $\Brho_K$ on the SINR with i.i.d. channels. The system dimensions are set as $N = 72$ and $M = 40$. The parameter $\Brho_K$ is of the form $\Brho_K = [\frac{\rho}{N_1}, \ldots, \frac{\rho}{N_K}]^T$.  It can be observed that SINR is not sensitive to $\rho$. Significant performance loss is only observed when $\rho$ is very large. Furthermore, according to Corollary \ref{Coro_best_rho}, the theoretical optimal parameter $\Brho_K^{\mathrm{opt}}$ should be $\Brho_K^{\mathrm{opt}} = (-30 \mathrm{dB}) \cdot [\frac{1}{N_1}, \ldots, \frac{1}{N_1}]^T$. The optimal value $\rho^{\mathrm{opt}} = -30$ dB shown in the figure obtained by the exhausted search validates the accuracy of Corollary \ref{Coro_best_rho}.
\par
\begin{figure}[t]
\centering
\vspace{-2mm}
\includegraphics[width=3.1in]{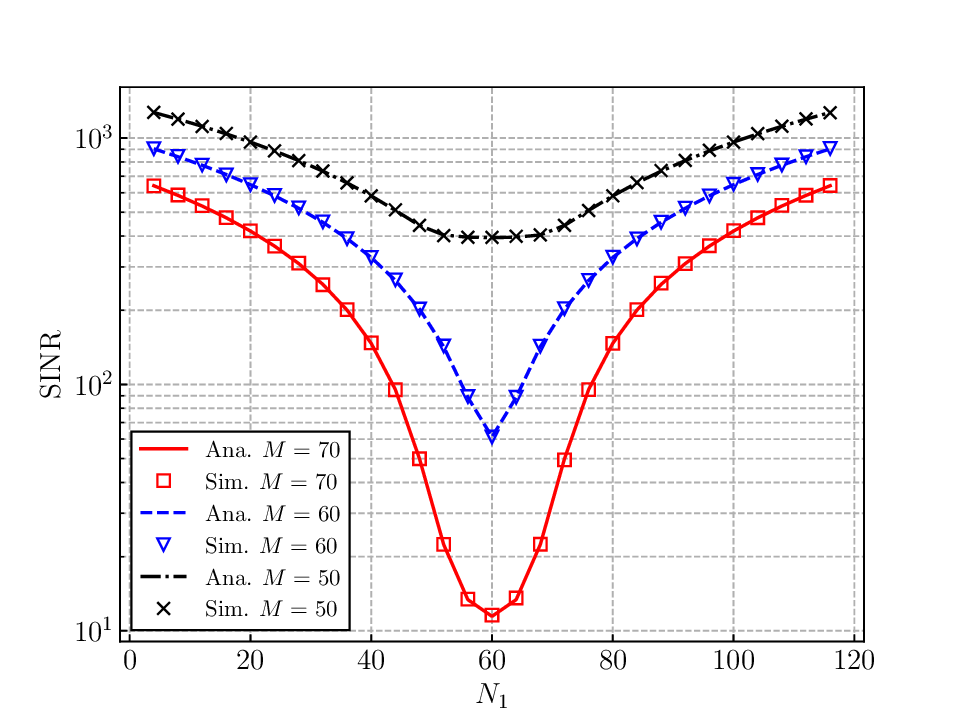}
\vspace{-2mm}
\caption{SINR versus the number of antennas allocated to the first cluster $N_1$.} 
\label{Figiid_2}
\end{figure}

\textbf{Impact of Antenna Partitioning Strategy:} In Fig. \ref{Figiid_2}, we plot the SINR for different antenna partitioning strategies with different numbers of users. The parameters are set as $N = 120$, $K = 2$, and $\rho_k = \frac{\sigma^2}{N_k}$. It is observed that when clusters are of equal size, the SINR is the lowest, which agrees with Corollary \ref{Coro_best_c}. Furthermore, as the number of users increases, the performance degradation with uniform antenna partitioning becomes more obvious.
\par
\begin{figure}[t]
\centering
\includegraphics[width=3.1in]{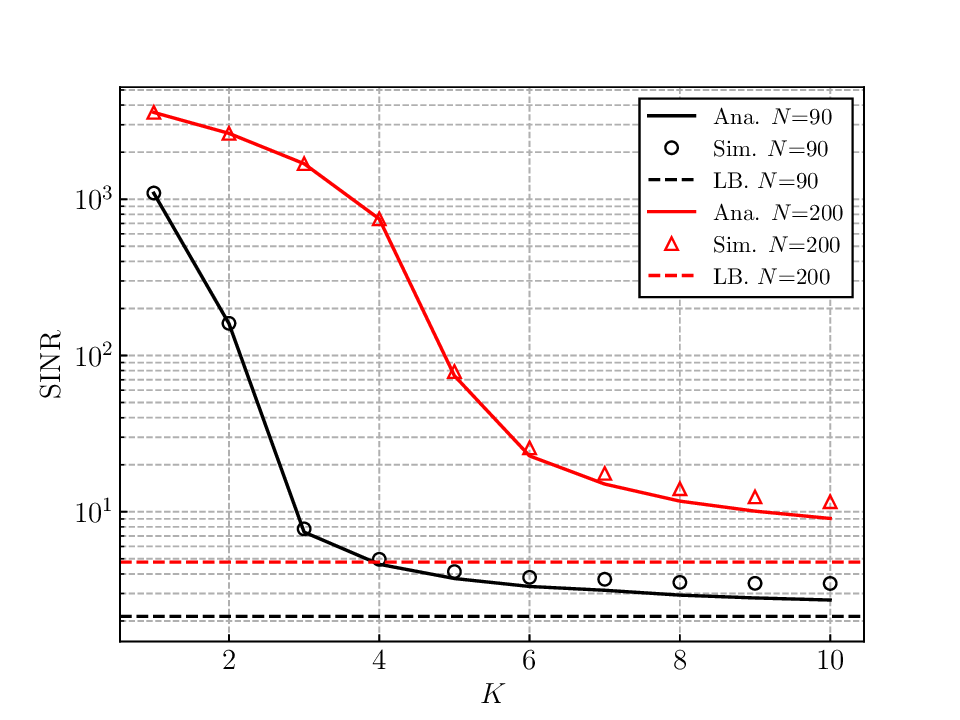}
\vspace*{-2mm}
\caption{SINR versus $K$.} 
\label{Figiid_3}
\end{figure}

\textbf{Impact of Number of Clusters:} 
 In Fig. \ref{Figiid_3}, we plot the SINR for different numbers of clusters $K$ where antennas are almost equally allocated with $\mathcal{N}_K = (\lfloor\frac{N}{K}\rfloor, \lfloor\frac{N}{K}\rfloor, \ldots, N - (K-1)\lfloor\frac{N}{K} \rfloor)$. The parameters are set as $M = 40$ and $\rho_k = \frac{\sigma^2}{N_k}$.  It can be observed that as the number of clusters increases, the SINR decreases monotonically and is lower bounded by \eqref{Eq_SINR_K_LowerBd}, which is shown by the dashed lines. This validates Corollary \ref{Coro_impact_K}.
\vspace{-2mm}
\section{Conclusion}
\label{Sec_Conclu}
In this paper, we considered the decentralized MIMO systems and imperfect CSI using LMMSE receivers. It was shown that with DBP, the linear fusion coefficients that maximize the receive SINR also minimize the MSE for estimating the transmit signal. To reduce the complexity, we further proposed two linear fusion schemes and derived the deterministic approximations of
the SINR with imperfect CSI and general spatial correlations. Based on the results, we investigated the optimal LMMSE regularization parameters, optimal antenna partitioning strategies, and the impact of the number of clusters. It was shown that with i.i.d. channels, the SINR is minimized when the clusters are of equal size. Conversely, the optimal performance is achieved when antennas are mostly unbalanced distributed, i.e., all antennas are allocated to a single cluster. We also proved that as the number of clusters increases, the SINR monotonically decreases. 
\par
From the perspective of RMT, we investigated the asymptotic convergence properties of the resolvent for the covariance matrix with generally correlated columns and its related functions. These results can be utilized for the second-order SINR analysis, e.g., the outage probability, and be applied to downlink scenarios, e.g., the analysis of RZF.

\appendices
\section{Useful Results}
\label{App_Useful_Results}
In this section, we will give the main mathematical results, which will be frequently used throughout the proofs in this paper.
\vspace*{-3mm}
\subsection{Matrix Inversion Lemma}
\begin{lemma} \cite[Lemma 1]{Wagner2012MISO}: 
\label{Lem_Wood_matrix}
Let $\BM \in \mathbb{C}^{N \times N}$ be an invertible matrix, $\Bm \in \mathbb{C}^N$, and $a \in \mathbb{C}$. If $1 + a \Bm^H \BM^{-1}\Bm \neq 0$, we have
\begin{equation}
    \Bm^H\left( \BM + a \Bm\Bm^H \right)^{-1} = \frac{\Bm^H\BM^{-1}}{1 + a \Bm^H \BM^{-1}\Bm}.
\end{equation}
\end{lemma}
\vspace*{-5mm}
\subsection{Convergence of Random Vectors}
\begin{lemma}
\label{Lemma_Trace_as}
\cite[Lemma 4]{Wagner2012MISO}:    Let  $\BA_N \in \mathbb{C}^{N \times N}$ be a random matrix and  $\Bx_N \in \mathbb{C}^N$ be a random vector with i.i.d. entries with normalized Gaussian distribution $\mathcal{CN}(0, 1)$. Assume that for each $j$, $[\Bx]_j$ is independent with sequence $(\BA_N)_{N \geq 1}$ and $\mathbb{P}(\sup_{N \geq 1} \norm{\BA_N} < + \infty) = 1$. Then, we have
    \begin{equation}
        \frac{\Tr (\Bx^H_N \BA_N \Bx_N)}{N}  - \frac{\Tr \BA_N}{N} \xrightarrow[N \rightarrow + \infty]{a.s.} 0.
    \end{equation}
\end{lemma}
\textit{Proof:} Here we provide a proof different from \cite{Wagner2012MISO} which relies on Tonelli's theorem. To describe independence, \cite{Wagner2012MISO} assumes the underlying probability space is the product probability space of $(\Bx_N)_{N\geq 1}$ and $(\BA_N)_{N\geq1}$, whereas here we discuss within the same probability space. 
\par
Denoting the underline probability space $(\Omega, \mathcal{F}, \mathbb{P})$. Define the set $A_{n, K} = \{\omega \in \Omega:  \sup_{j \geq n} \norm{\BA_j(\omega)} \leq K\}$ and $B_{N, \epsilon}= \{\omega \in \Omega: d_N(\omega) \leq \epsilon \}$, where 
\begin{equation}
    d_N(\omega) = \abs{\frac{\Tr [\Bx^H_N(\omega) \BA_N(\omega) \Bx_N(\omega)] }{N}  - \frac{\Tr \BA_N(\omega)}{N}}.
\end{equation}
To prove $d_N \xrightarrow[N \rightarrow + \infty]{a.s.}$ 0, we only need to show that for any positive integer $n$ and $\epsilon > 0$, $\mathbb{P}(\bigcup_{n=1}^{+ \infty}\bigcap_{N=n}^{+ \infty}  B_{N, \epsilon}) = 1$ holds. In the following, we will prove this by a stronger result $\mathbb{P}(\bigcup_{n=1}^{+ \infty}\bigcap_{N=n}^{+ \infty}  B_{N, \epsilon}) \geq 1$. To this end, we write
\begin{equation}
\begin{split}
\label{Eq_as}
    &\mathbb{P}( \bigcup_{n=1}^{+ \infty}\bigcap_{N=n}^{+ \infty}  B_{N, \epsilon} ) \geq \mathbb{P}(\bigcup_{n=1}^{+ \infty} \bigcap_{N=n}^{+ \infty}  B_{N, \epsilon} \bigcap\bigcup_{K=0}^{+ \infty}A_{n,K}) \\
    & = \mathbb{P}( \bigcup_{K=0}^{+ \infty} \bigcup_{n=1}^{+ \infty}\bigcap_{N=n}^{+ \infty}  B_{N, \epsilon} \bigcap A_{n,K}) \\
    &\overset{(a)}{=} \lim_{K \rightarrow + \infty} \lim_{n \rightarrow + \infty} \mathbb{P}( \bigcap_{N=n}^{+ \infty}  B_{N, \epsilon} \bigcap A_{n,K}),
\end{split}
\end{equation}
where step $(a)$ follows from $ A_{n,K} \subset A_{n, K+1}$,  $ A_{n,K} \subset A_{n + 1, K}$, and  continuity of probability measure. On the other side, we have
\begin{equation}
\begin{split}
    &\mathbb{P}( \bigcap_{N=n}^{+ \infty}  B_{N, \epsilon} \bigcap A_{n,K} ) = 1 - \mathbb{P}( \bigcup_{N=n}^{+ \infty}  B_{N, \epsilon}^c \bigcup A_{n,K}^c) \\
    & = 1 - \mathbb{P}(A_{n,K}^c) - \mathbb{P}( \bigcup_{N=n}^{+ \infty}  B_{N, \epsilon}^c \bigcap A_{n,K}) \\
    & \geq \mathbb{P}(A_{n,K}) - \sum_{N \geq n} \mathbb{P}( B_{N, \epsilon}^c \bigcap A_{n,K}).
\end{split}
\end{equation}
By using Markov's inequality, we can get
\begin{equation}
    \begin{split}
        & \mathbb{P}( B_{N, \epsilon}^c \bigcap A_{n,K} )  \leq \E \mathbb{I}_{\{ d_N \geq \epsilon \}} \mathbb{I}_{\{\norm{\BA_N} \leq K\}} \\
        &\leq \E \left[ \mathbb{I}_{\{\norm{\BA_N} \leq K\}} \E ( \epsilon^{-p}{d_N^p}| \BA_N )  \right] \\
        &\overset{(a)}{\leq} C_p N^{-\frac{p}{2}} \E [ \mathbb{I}_{\{\norm{\BA_N} \leq K\}} \norm{\BA_N}^p] = \mathcal{O}(N^{-\frac{p}{2}}{K^p}),
    \end{split}
    \end{equation}
where $p > 2$, $C_p$ is a constant, and step $(a)$ follows from the trace lemma \cite[Lemma 3]{Wagner2012MISO}. Hence, $\sum_{N = n}^{+ \infty } \mathbb{P} (B_{N, \epsilon}^c \bigcap A_{n,K}) \rightarrow 0$ as $n \rightarrow + \infty$ and it holds true that 
\begin{equation}
\label{Eq_ninf}
    \lim_{n \rightarrow + \infty} \mathbb{P}( \bigcap_{N=n}^{+ \infty}  B_{N, \epsilon} \bigcap A_{n,K}) \geq \lim_{n \rightarrow + \infty} \mathbb{P}(A_{n,K}). 
\end{equation}
By combining \eqref{Eq_ninf} and \eqref{Eq_as}, we can obtain 
\begin{equation}
    \begin{split}
        &\mathbb{P}( \bigcup_{n=1}^{+ \infty}\bigcap_{N=n}^{+ \infty}  B_{N, \epsilon}) {\geq} \lim_{K \rightarrow + \infty} \lim_{n \rightarrow + \infty}\mathbb{P}(A_{n,K}) \\
        & \overset{(a)}{=}  \mathbb{P}(\bigcup_{k=0}^{+ \infty}\bigcup_{n=1}^{+ \infty} A_{n, K}) =  \mathbb{P}(\bigcup_{n=1}^{+ \infty}\bigcup_{K=0}^{+ \infty} A_{n, K}) \overset{(b)}{=} 1,
    \end{split}
\end{equation}
where step $(a)$ follows from continuity of probability measure and step $(b)$ follows from the identity $\mathbb{P}(\bigcup_{K=0}^{+ \infty} A_{n, k}) = 1$. This completes the proof of Lemma \ref{Lemma_Trace_as}. \qed
\subsection{Convergence of Random Matrices}
The SINR involves terms associated with random matrices $\widehat{\BSigma}$ and $\widetilde{\BSigma}$ whose columns are generally correlated. This makes the evaluation of the SINR very challenging. In the following, we introduce a more general setting and provide the relevant results, which can be utilized to prove Theorem \ref{Thm_Asym_SINR_opt}.
\par
 Let $\widetilde{\BX} = [\BA_1\Bx_1,\cdots,  \BA_M\Bx_M] \in \C^{N \times M}$ and $\widetilde{\BY} = [\BB_1\Bx_1, \cdots, \BB_M\Bx_M] \in \C^{N \times M}$, where $\mathbf{A}_j, \BB_j \in \mathbb{C}^{N \times N_i^x}$, and  $\Bx_j \sim \mathcal{CN}(0,  \BI_{N^{x}_{j}})$ for each $j \in [M]$. Define two correlation matrices sets
$A = \{\BA_j: j\in [M]\}$ and $B = \{\BB_j: j\in [M]\}$. Denote $\BOmega_j = \BA_j \BA_j^H$, $\BA_{jk} = \BA_j[\Bvarsigma_k, :]$, and $\BB_{jk} = \BB_j[\Bvarsigma_k, :]$.  Let $\BX_k = \frac{1}{\sqrt{N}_k}\widetilde{\BX}[\Bvarsigma_k, :]$ and $\BY_k = \frac{1}{\sqrt{N_k}}\widetilde{\BY}[\Bvarsigma_k, :]$. The resolvent matrix of ${\BX_k\BX_k^H} + \BS_k$ is given by
 \begin{equation}
     \BQ_k(z) = \left( {\BX_k\BX_k^H} + \BS_k - z \BI_{N_k}\right)^{-1},
 \end{equation}
 with $\BS_k $ being deterministic and Hermitian nonnegative. We further define some functions of the resolvents in Table \ref{Table_of_notations}, where we set $\BQ_k = \BQ_k(z_k)$, for each $k \in [K]$ and omit the variable $z_k$ for the sake of notation simplicity. Furthermore, $\BBA$ and $\BBB$ are appropriately sized matrices, and $\Bb$ is a properly dimensioned vector. 
 \begin{table}[t]
    \centering
    \caption{Table of Notations}
    \vspace{-2mm}
    \begin{tabular}{|c|c|}
    \toprule[1pt]
        Notation & Expression \\
        \midrule
        $\digamma_{k}(\BBA)$ & $\Tr \BBA \BQ_k$
        \\ 
        $\Phi_{B, kl} (\BBA, \Bb)$ & $\Tr \BBA \BQ_k \BX_k \diag(\Bb) \BY_l^H$\\
        [0.5ex]
        $\Phi_{kl}(\BBA, \Bb)$ & $\Tr \BBA \BQ_k \BX_k\diag(\Bb) \BX_l^H$ \\
        [0.5ex]
        $ \Upsilon_{kl}( \BBA, \BBB )$ & $\Tr  \BBA \BQ_k  \BBB \BQ_l$\\
        [0.5ex]
        $  \Pi_{B, kl}(\BBA)$ & $\Tr  \BBA \BQ_k \BY_k \BY_l^H \BQ_l$ \\
        [0.5ex]
        $\Pi_{kl}(\BBA)$ & $\Tr  \BBA \BQ_k \BX_k \BX_l^H \BQ_l$ \\
        [0.5ex]
    \bottomrule[1pt]
    \end{tabular}
    \label{Table_of_notations}
\end{table}
We next introduce the stochastic equivalent notation. 
 \begin{definition} (Stochastic Equivalent)
      For random sequence $(X_N)_{N \geq 1}$ and deterministic sequence $(\overline{X}_N)_{N \geq 1}$, we use $X_N {\asymp}_{l} \overline{X}_N$ to denote
 \begin{equation}
    \abs{\E { X}_N - \overline{X}_N} = \BO(N^{-l}), ~~ \frac{\abs{X_N - \overline{X}_N}}{N} \xrightarrow[N \rightarrow + \infty]{a.s.} 0. 
\end{equation}
 \end{definition}
The following lemma gives the stochastic equivalents for the functions in Table \ref{Table_of_notations}.
\begin{lemma}
\label{Thm_DE}
    Assume Assumption \textbf{\ref{A.1}} holds, the maximal spectral norm of the correlation matrices is bounded, i.e., $\limsup_{N} \max_{j \in [M]} \{\norm{\BA_j}, \norm{\BB_j} \} < + \infty$, and the spectral norms of matrices $\BBA, \BBB$ and $\diag(\Bb)$ are uniformly bounded. For any $z_1, \ldots, z_K < 0$, we have 
    $\digamma_k \asymp_{\frac{1}{2}} \overline{\digamma}_k$, $\Phi_{B, kl}  \asymp_{\frac{1}{2}} \overline{\Phi}_{B, kl}$, $\Phi_{kl}  \asymp_{\frac{1}{2}} \overline{\Phi}_{kl}$, $\Upsilon_{kl}\asymp_{\frac{1}{2}} \overline{\Upsilon}_{kl}$, ${\Pi}_{B, kl} \asymp_{\frac{1}{2}} \overline{\Pi}_{B, kl}$, and  ${\Pi}_{kl} \asymp_{\frac{1}{2}} \overline{\Pi}_{kl}$, where
    \begin{align}
    \overline{\digamma}_k (\BBA) &=  \Tr \BBA \BTheta_k, \label{DE_digamma}\\
          \overline{\Phi}_{B, kl}(\BBA, \Bb) &=  \frac{\sum_{j=1}^M [ \widetilde{\BF}_k\Bb]_j \Tr \BBA \BTheta_k \BA_{jk}\BB_{jl}^H }{\sqrt{N_kN_l}} , \label{DE_Phi}\\
         \overline{\Upsilon}_{kl}( \BBA, \BBB ) &=  \Tr \BBA \BTheta_k \BBB \BTheta_l \notag \\
         &+  \widetilde{\Blambda}_{kl}( \BBA) \widetilde{\BF}_{k} \widetilde{\BF}_{l} {\BXi}_{kl}{\Blambda}_{kl}(\BBB), \label{DE_Upsilon}
    \end{align}
     and $\overline{\Pi}_{kl, B}$ is given at the top of the next page. Functions $\overline{\Phi}_{kl}$ and $\overline{\Pi}_{kl}$ can be obtained by setting $\BB_j$ = $\BA_j$ for each $j \in [M]$ in $\overline{\Phi}_{kl, B}$ and $\overline{\Pi}_{kl, B}$, respectively and we omit them here. The deterministic variables are defined in Table \ref{Table_Deter_Terms} where $P $ and $ O$ can be any symbol from the set $\{ A, B \}$. The matrix $\BTheta_k$ is defined by the positive solutions of the following system of equations w.r.t. $\delta_{jk}$
     \setcounter{equation}{56}
    \begin{align}
        \delta_{jk} &= \frac{1}{N_k} \Tr [\BOmega_{j}]_{[k, k]} \BTheta_k, j \in [M], \\
        \BTheta_k &= \left(-z_k\BI_{N_k} + \BS_k + \frac{1}{N_k}\sum_{j=1}^M \frac{[\BOmega_{j}]_{[k, k]} }{1 + \delta_{jk}} \right)^{-1}.
    \end{align}  
\end{lemma}
\begin{table*}[t]
    \centering
    \caption{Table of Deterministic Quantities\vspace*{-2mm}}
    \label{Table_Deter_Terms}
    \begin{tabular}{|cc|cc|}
    \toprule[1pt]
    Notation &  Expression  & Notation &  Expression\\
    \midrule
        $\widetilde{\BF}_k$ & $\diag \left(\left(1 + \delta_{jk}
    \right)^{-1}; j \in [M] \right)$ &
        $[\BGamma_{kl}]_{i, j}$ & $\frac{1}{N_kN_l} \Tr [\BOmega_{i}]_{[l, k]} \BTheta_k [\BOmega_{j}]_{[k, l]} \BTheta_l$ \\
        $\BXi_{kl}$ & $\left(\BI_M - \BGamma_{kl} \widetilde{\BF}_{k} \widetilde{\BF}_{l}  \right)^{-1}$ & $ \widetilde{\Blambda}_{kl}( \BBA)$ & $\frac{\mathbf{1}_M^T}{\sqrt{N_kN_l}}\diag\left(\Tr\BBA \BTheta_k [\BOmega_j]_{[k, l]} \BTheta_l; j\in [M] \right)$ \\
        $\widetilde{\Blambda}_{P, O, kl}(\BBA)$ &$\frac{\mathbf{1}_M^T}{\sqrt{N_kN_l}} \diag\left( \Tr\BBA \BTheta_k \BP_{jk} \mathbf{O}_{jl}^H \BTheta_{l}; j \in [M] \right)$& ${\Blambda}_{kl}(\BBB)$& $\frac{1}{\sqrt{N_kN_l}}\diag\left(\Tr[\BOmega_j]_{[l, k]} \BTheta_k\BBB \BTheta_l; j\in [M] \right) \mathbf{1}_M$ \\
        $[\BLambda_{P, O, kl}]_{i, j}$&
        $ \frac{1}{N_kN_l}\Tr [\BOmega_i]_{[l, k]} \BTheta_k\BP_{jk}\mathbf{O}_{jl}^H \BTheta_l$
        &$ \BD_{P, O, k}$& $\frac{1}{N_k} \diag\left( \Tr \BP_{jk} \mathbf{O}_{jk}^H \BTheta_k; j \in [M] \right)$ \\
        \bottomrule[1pt]
    \end{tabular}
\end{table*}

\begin{figure*}[!t]
\vspace*{-4mm}
\setcounter{equation}{55}
\begin{align}
     \overline{\Pi}_{kl, B}(\BBA ) &=  \left[ \widetilde{\Blambda}_{B, B, kl}(  \BBA) - \widetilde{\Blambda}_{B, A, kl}(  \BBA)\BD_{A, B, l} \widetilde{\BF}_{l} \notag - \widetilde{\Blambda}_{A, B, kl}( \BBA)\BD_{B, A, k} \widetilde{\BF}_{k} \right]\mathbf{1}_M   \\
    &+   \widetilde{\Blambda}_{kl}( \BBA) \widetilde{\BF}_{k} \widetilde{\BF}_{l} {\BXi}_{kl} \left[ {\BLambda}_{B, B, kl}  - {\BLambda}_{B, A, kl}\BD_{A, B, l} \widetilde{\BF}_{l} - {\BLambda}_{A, B, kl}\BD_{ B, A, k} \widetilde{\BF}_{k} + \BD_{B, A, k} \BD_{A, B, l} \right] \mathbf{1}_M.
\end{align}
\hrulefill
\vspace{-6mm}
\end{figure*}
\textit{Proof:} The proof of Lemma \ref{Thm_DE} is given in Appendix \ref{App_Thm_DE}. \qed

\section{Proof of Proposition \ref{Prop_opt_alpha}}
\label{APP_Prop_opt_alpha}
Denote $\widetilde{\BH} = [\widetilde{\Bh}_1, \ldots, \widetilde{\Bh}_M]$ and $\widehat{\BH}_k = [\widehat{\Bh}_{1k}, \ldots, \widehat{\Bh}_{Mk}]$. By taking $\Br_k = \Br_k^{\mathrm{mmse}}$ in \eqref{SINR_Exp}, straight-forward computations yield
\setcounter{equation}{58}
\begin{align}
\label{EQ_gamma_f}
    \gamma_0 = \frac{\Balpha \Bm \Bm^H \Balpha^H }{ \Balpha \BM \Balpha^H },
\end{align}
where $ \BM = \BD_{r}^H (  \widetilde{\BH}\widetilde{\BH}^H + \BW + \sigma^2 \BI_N) \BD_{r}$ and $\Bm = \BD_{r}^H\widetilde{\Bh}_0 $. One can validate that $\BM$ is almost surely invertible. By letting $\widetilde{\Balpha} =  \Balpha \BM^{\frac{1}{2}}$ and $\widetilde{\Bm} = \BM^{-\frac{1}{2}} \Bm $, we have
\begin{equation}
\label{Eq_SINR_ratio_2}
    \gamma_0 = \frac{\widetilde{\Balpha} \widetilde{\Bm}\widetilde{\Bm}^H\widetilde{\Balpha}^H}{\widetilde{\Balpha}\widetilde{\Balpha}^H},
\end{equation}
where $\gamma_0$ is maximized when $\widetilde{\Balpha} = \widetilde{c} \widetilde{\Bm}^H$ for 
$\widetilde{c} \neq 0$ according to Cauchy-Schwartz inequality.  The optimal $\Balpha$ is given by
\begin{equation}
\begin{split}
    \Balpha^{\mathrm{opt}} &= \widetilde{c} \Bm^H \BM^{-1}   =  \frac{ c \Bm^H\BM^{-1} }{1 + \Bm^H \BM^{-1} \Bm } \\
    &\overset{(a)}{=} c \Bm^H\left(\BM + \Bm\Bm^H \right)^{-1},
\end{split}
\end{equation}
where $c = \widetilde{c} (1 + \Bm^H \BM^{-1} \Bm)$ and step $(a)$ follows from Lemma \ref{Lem_Wood_matrix}. Thus we proved \eqref{Eq_opt_alpha}. Since \eqref{Eq_opt_alpha_MMSE} is convex, by setting the derivative of the optimization objective w.r.t. $\Balpha$ to be to 0, we can obtain that $\Balpha^{\mathrm{opt}}$ is the optimal solution when $c = 1$. Therefore, we complete the proof of Proposition \ref{Prop_opt_alpha}. \qed

\section{Proof of Theorem \ref{Thm_Asym_SINR_opt}}
\label{APP_Thm_Asym_SINR_opt}
\subsection{Proof of the Almost Sure Convergence}
According to Lemma \ref{Lem_Wood_matrix}, we can write 
\begin{equation}
\label{Eq_Dr}
    \BD_{r} = \BQ \BD_{h, 0} \BL \BF,
\end{equation}
where
\begin{align}
    \BQ &= \diag (\BQ_k; k\in [K]), ~ \BQ_k = \Big(\frac{\widehat{\BH}_k\widehat{\BH}_k^H}{N_k} + \BZ_k + \rho_k \BI_{N_k} \Big)^{-1}, \notag\\
    \BL &= \diag(N_k^{-1}; k \in [K]), ~ \BD_{h, 0} = \diag(\widehat{\Bh}_{0k};k\in [K]) \notag \\
    \BF &= \diag((1 + \frac{1}{N_k}{\widehat{\Bh}}_{0k}^H\BQ_k\widehat{\Bh}_{0k})^{-1}; k\in[K]).
\end{align}
We first analyze the LFSC case.
By setting $a=1$, $\Bm = \BD_{r}^H \widehat{\Bh}_0$, and $\BM = \BD_{r}^H (  \widehat{\BH}\widehat{\BH}^H + \BD_{W} + \sigma^2 \BI_N) \BD_{r}$ in Lemma \ref{Lem_Wood_matrix} and using \eqref{Eq_Dr}, we can obtain
\begin{equation}
    \gamma^{\mathrm{lfsc}} = \frac{\abs{\Bg^H \BG_{\Delta_I}^{-1} \widetilde{\Bg} }^2}{\Bg^H \BG_{\Delta_I}^{-1}\BG_{\Delta}\BG_{\Delta_I}^{-1} \Bg},
\end{equation}
where
\begin{align}
    \Bg^H &= \widehat{\Bh}_0^H\BQ\BD_{h, 0} \BL, ~~ \widetilde{\Bg}^H = \widetilde{\Bh}_0^H\BQ\BD_{h, 0} \BL,\notag \\
    \BG_{\Delta_I} &= \BL\BD_{h, 0}^H \BQ (\widehat{\BH}\widehat{\BH}^H + \BD_{W} + \sigma^2 \BI_N) \BQ \BD_{h, 0} \BL, \notag \\
    \BG_{\Delta} &= \BL\BD_{h, 0}^H \BQ (\widetilde{\BH}\widetilde{\BH}^H + \BW + \sigma^2 \BI_N) \BQ \BD_{h, 0} \BL.
\end{align}
It can be proved that $\norm{\BQ} = \sup_k \norm{\BQ_k} \leq \frac{1}{\min_k \rho_k}$ for $\rho_k > 0$. According to Lemma \ref{Lemma_Trace_as} and \ref{Thm_DE}, we have
\begin{equation}
    \begin{split}
        &[\Bg]_k - \frac{\Tr [\BPhi_{0}]_{[k, k]} \BTheta_k}{N_k}  = \frac{ \widehat{\Bh}_{0k}^H \BQ_k \widehat{\Bh}_{0k}}{N_k} - \frac{\Tr [\BPhi_{0}]_{[k, k]} \BTheta_k}{N_k} \\
        & =   \frac{ {\Bq}_{0}^H (\BPhi^{\frac{1}{2}}_{0k})^H\BQ_k \BPhi^{\frac{1}{2}}_{0k}{\Bq}_{0}}{N_k} - \frac{\mathrm{Tr} (\BPhi^{\frac{1}{2}}_{0k} )^H \BQ_k \BPhi^{\frac{1}{2}}_{0k}}{N_k} \\
        &+ \frac{\Tr [\BPhi_0]_{[k, k]}\BQ_k }{N_k} -  \frac{\Tr [\BPhi_{0}]_{[k, k]} \BTheta_k}{N_k}  \xrightarrow[N \xrightarrow{c_1, \ldots, c_K} + \infty]{ a.s.}  0,
    \end{split}
\end{equation}
where ${\Bq}_{0} \sim \mathcal{CN}(0, \BI_N) $ and $\BPhi_{0k}^{\frac{1}{2}} = \BPhi_0^{\frac{1}{2}}[\Bvarsigma_k, :] \in \mathbb{C}^{N_k \times N}$. Since $K$ is a given number, it holds true that $[\Bg]_k - [\Bv]_k$ converges to 0 almost surely for each $k \in [K]$. Similarly, we write 
\begin{equation}
\begin{split}
    [\BG_{\Delta}]_{k, l} &=   \frac{\widehat{\Bh}_{0k}^H\BQ_k([\BW]_{[k, l]} + \sigma^2 \mathbb{I}_{\{k=l\}}\BI_{N_k})\BQ_l\widehat{\Bh}_{0l}}{N_kN_l} \\
    &+ \frac{\widehat{\Bh}_{0k}^H\BQ_k \widetilde{\BH}_{k} \widetilde{\BH}_l^H\BQ_l\widehat{\Bh}_{0l}}{N_kN_l}  = G_{kl, 1} + G_{kl, 2}. 
\end{split}
\end{equation}
According to \cite[Theorem 2]{Abla2016Noeigen}, we can conclude that the spectral norm of matrix $\frac{1}{\sqrt{N_k}}\widetilde{\BH}_k$ is bounded with probability $1$ for each $k \in [K]$ such that the following holds almost surely 
\begin{equation}
\begin{split}
    &\sup_{N \geq 1} \lVert \frac{\BQ_k \widetilde{\BH}_{k} \widetilde{\BH}_l^H\BQ_l}{\sqrt{N_kN_l}} \rVert \leq \sup_{N \geq 1} \frac{\lVert \frac{ \widetilde{\BH}_{k}} {\sqrt{N_k}}\rVert \lVert \frac{ \widetilde{\BH}_l} {\sqrt{N_l}}\rVert }{\min_k \rho_k^2} < + \infty,
\end{split}
\end{equation}
for each $k, l \in [K]$. By Lemma \ref{Lemma_Trace_as} and \ref{Thm_DE}, the difference $ \mathcal{D}_{G_{kl, 2}} =  G_{kl, 2} - \frac{1}{\sqrt{N_kN_l}}{\overline{\Pi}}_{B, kl}([\BPhi_0]_{[l, k]})$ can be written as
\begin{equation}
\begin{split}
    &\mathcal{D}_{G_{kl, 2}}  = G_{kl, 2} - \frac{\Tr  \BQ_k\widetilde{\BH}_{k} \widetilde{\BH}_l^H\BQ_l [\BPhi_0]_{[l, k]} }{N_kN_l}  \\
    &~~~ + \frac{\Tr  \BQ_k\widetilde{\BH}_{k} \widetilde{\BH}_l^H\BQ_l [\BPhi_0]_{[l, k]} }{N_kN_l} - \frac{{\overline{\Pi}}_{B, kl}([\BPhi_0]_{[l, k]}) }{\sqrt{N_kN_l}},
\end{split}
\end{equation}
which converges to 0 almost surely. Following a similar derivation, we can prove $G_{kl, 1}$ converges almost surely to $ \frac{1}{N_kN_l}\overline{\Upsilon}_{kl}( [\BPhi_{0}]_{[l, k]}, [\BW + \sigma^2 \BI_N]_{[k, l]})$. Moreover, it can be shown that, 
\begin{align}
    [\widetilde{\Bg}]_k - [\Bv]_k &\xrightarrow[N \xrightarrow{c_1, \ldots, c_K} + \infty]{ a.s.} 0, 
    \\
     [\BG_{\Delta_I}]_{k, l} - [\BDelta_I]_{k, l} &\xrightarrow[N \xrightarrow{c_1, \ldots, c_K} + \infty]{ a.s.} 0,
 \end{align}
for each $k, l \in [K]$. According to the continuous mapping theorem \cite{van2000asymptotic}, we have 
\begin{align}
    \gamma^{\mathrm{lfsc}}  \xrightarrow[N \xrightarrow{c_1, \ldots, c_K} + \infty]{ a.s.} \overline{\gamma}^{\mathrm{lfsc}},
\end{align}
which concludes \eqref{Eq_SINR_LFSC}. The proof for \eqref{Eq_SINR_LFOC} an \eqref{Eq_SINR_LFCC} can be obtained similarly and is omitted here. \qed
{ 
\subsection{Proof of the Approximation for the Mean}
We analyze the LFOC case here and the other two cases are similar.  By using the resolvent identity $\BA - \BB = \BA(\BB^{-1} - \BA^{-1})\BB$, we have
\begin{equation}
\begin{split}
\label{Eq_G}
    \BG^{-1}_{\Delta} &= \BDelta^{-1} + \BDelta^{-1} (\BDelta - \BG_{\Delta}) \BDelta^{-1} \\
    &+ \BG^{-1}_{\Delta}(\BDelta - \BG_{\Delta}) \BDelta^{-1} (\BDelta - \BG_{\Delta}) \BDelta^{-1}.
\end{split}
\end{equation}
According to the definition, $\gamma^{\mathrm{lfoc}}$ can be rewritten as \eqref{Eq_gamma_sp} at the top of the next page.
\begin{figure*}[!t]
 
\begin{equation}
\begin{split}
\label{Eq_gamma_sp}
    \gamma^{\mathrm{lfoc}} &= \Bv^H \BDelta^{-1} \Bv + \Bv^H \BDelta^{-1} (\widetilde{\Bg} - \Bv)+(\widetilde{\Bg} - \Bv)^H \BDelta^{-1} \Bv + (\widetilde{\Bg} - \Bv)^H \BDelta^{-1} (\widetilde{\Bg} - \Bv)
        + \Bv^H\BDelta^{-1}(\BDelta - \BG_{\Delta})\BDelta^{-1}\Bv \\
        &+  \Bv^H \BDelta^{-1}(\BDelta - \BG_{\Delta})\BDelta^{-1} (\widetilde{\Bg} - \Bv)
        + (\widetilde{\Bg} - \Bv)^H \BDelta^{-1}(\BDelta - \BG_{\Delta})\BDelta^{-1} \Bv + (\widetilde{\Bg} - \Bv)^H \BDelta^{-1}(\BDelta - \BG_{\Delta})\BDelta^{-1} (\widetilde{\Bg} - \Bv) \\
        &+ \widetilde{\Bg}^H \BG^{-1}_{\Delta}(\BDelta - \BG_{\Delta}) \BDelta^{-1} (\BDelta - \BG_{\Delta}) \BDelta^{-1} \widetilde{\Bg} := \sum_{j=1}^9 X_j.
    \end{split}
    \end{equation}
\hrulefill
\vspace{-4mm}
\end{figure*}
\begin{figure*}[!t]
 
\setcounter{equation}{80}
\begin{equation}
\begin{split}
\label{Eq_EX6}
\abs{\E X_6} &= \abs{\E \Bv^H \BDelta^{-1}(\BDelta - \BG_{\Delta})\BDelta^{-1} (\widetilde{\Bg} - \Bv)} = |\E \sum_{i_1, i_2, i_3, i_4 \in [K]} [\Bv^H \BDelta^{-1}]_{i_1}[\BDelta^{-1}]_{i_1, i_2}  [\BDelta - \BG_{\Delta}]_{i_2, i_3}  [\BDelta^{-1}]_{i_3, i_4} [\widetilde{\Bg} - \Bv]_{i_4}| \\
& \overset{(a)}{\leq}  \sum_{i_1, i_2, i_3, i_4 \in [K]}  \abs{[\Bv^H \BDelta^{-1}]_{i_1}[\BDelta^{-1}]_{i_1, i_2}   [\BDelta^{-1}]_{i_3, i_4}} \E^{\frac{1}{2}}\abs{ [\BDelta - \BG_{\Delta}]_{i_2, i_3} }^2 \E^{\frac{1}{2}} \abs{ [\widetilde{\Bg}- \Bv]_{i_4}}^2   = \mathcal{O}(\frac{1}{N}).
\end{split}
\end{equation}
\hrulefill
\vspace{-4mm}
\end{figure*}
It can be proved that $\abs{\E X_j}$, $j = 2, 3, \ldots 9$ is of order $\mathcal{O}(N^{-1})$ and we will evaluate $|\E X_6|$ in the following, as the remaining terms can be addressed using a similar method. To this end, we first prove the following bounds
\setcounter{equation}{74}
\begin{align}
    &\E \abs{[\widetilde{\Bg} - \Bv]_k}^2 = \mathcal{O}(\frac{1}{N}), \forall k \in [K] \label{Eq_Variance_Control}\\
    &\E  \abs{[{\BG}_{\Delta} - \BDelta]_{k, l}}^2 = \mathcal{O}(\frac{1}{N}), \forall k, l \in [K]. \label{Eq_Variance_Control_1}
\end{align}
 Denote $\mathscr{F}_M$ as the $\sigma$-algebra generated by $\Bh_1, \ldots, \Bh_M$ and 
\begin{equation}
\underline{\Bg}_k  = \E ([\widetilde{\Bg}]_k| \mathscr{F}_M) = \frac{1}{N_k}\Tr [\BPhi_0]_{[k, k]} \BQ_k.
\end{equation}
According to the triangle inequality, we have
\begin{equation}
\label{Eq_bound_g}
\begin{split}
    \E |{[\widetilde{\Bg} - \Bv]_k}|^2 &\leq C (\E |[\widetilde{\Bg}]_k - \underline{\Bg}_k|^2 + \E |\underline{\Bg}_k - \E[\underline{\Bg}]_k|^2 \\
    &+ \abs{\E[\underline{\Bg}]_k - [\Bv]_k}^2) := C(W_1 + W_2 + W_3),
\end{split}
\end{equation}
where $C$ is a constant independent of $N$.
By applying the trace lemma \cite[Lemma 2.7]{bai1998no}, we have 
\begin{equation}
\begin{split}
W_1 &= \frac{\E \left[\E \left( \abs{\widehat{\Bh}_{0k}^H \BQ_k \widetilde{\Bh}_{0k} - \Tr\BQ_k [\BPhi_0]_{[k, k]}}^2  \big| \mathscr{F}_M \right) \right]}{N_k^2} \\
&\leq \frac{C_1}{N_k} \E \norm{\BQ_k}^2 = \mathcal{\BO}(\frac{1}{N}), 
\end{split}
\end{equation}
where $C_1$ is a constant. By using Poincaré-Nash inequality \cite[Eq. (18)]{Hachem2008ANewApproach}, we can get 
\begin{equation}
W_2 = \mathrm{Var}(\underline{\Bg}_k) \leq \mathcal{O}(\frac{1}{N^2}).
\end{equation}
Finally, by Lemma \ref{Thm_DE}, we have $W_3 \leq \mathcal{O}(\frac{1}{N^3})$. Therefore, the dominant term on the RHS of \eqref{Eq_bound_g} is $\BO(N^{-1})$ which proves \eqref{Eq_Variance_Control}. The bound \eqref{Eq_Variance_Control_1} can be proved in a similar manner. 
Next, we evaluate the term $|\E X_6|$. By applying \eqref{Eq_Variance_Control} and \eqref{Eq_Variance_Control_1}, bound \eqref{Eq_EX6} at the top of the next page can be derived, where step $(a)$ follows from Cauchy-Schwartz inequality $\E|XY| \leq \E^{\frac{1}{2}} |X|^2  \E^{\frac{1}{2}} |Y|^2 $. 
Therefore, Theorem \ref{Thm_Asym_SINR_opt} is proved.  \qed}
{ 
\vspace{-1mm}
\setcounter{equation}{81}
\section{Proof of Corollary \ref{Coro_Unbalance_CSI}}
\label{App_Proof_Coro_Unbalance_CSI}
Define $\BR_j^{\mathcal{K}_g}$, $\BD_{\widetilde{\sigma}}^{\mathcal{K}_g}$, $\BD_{R, j}^{\mathcal{K}_g}$, $\BT_j^{\mathcal{K}_g}$, $\BD_{T, j}^{\mathcal{K}_g}$, $\BPhi_j^{\mathcal{K}_g}, \BV_j^{\mathcal{K}_g}$, $\BW_j^{\mathcal{K}_g}$, and $\BW^{\mathcal{K}_g}$ as the matrices obtained by considering the channel $\Bh^{\mathcal{K}_g}_j$ in \eqref{channel_disttribution} and \eqref{Eq_Def_Mat_CE} for each $j \in [M]_0$. 
Denote the asymptotic regime $\widetilde{\sigma}^2_e \rightarrow + \infty$ as $\widetilde{\sigma}^2_k \rightarrow + \infty$ for each $k \in \mathcal{K}_e$ with equal value $\widetilde{\sigma}_k^2 = \widetilde{\sigma}^2_e$. 
To prove Corollary \ref{Coro_Unbalance_CSI}, it is sufficient to show 
\begin{itemize}
    \item [1)] $[\BR^{\mathcal{K}_g}]_{[a, b]} \rightarrow [\BR]_{[k_a, k_b]}$, for $a, b \in [G]$ as $\widetilde{\sigma}^2_e \rightarrow + \infty$. Here, $\BR$ is one of
    \begin{equation}
    \left\{\BPhi_j, \BV_j\BPhi_j, \BV_j\BPhi_j\BV_j^H\right\}_{j=0}^M \cup \left\{ \BW, \BD_{W}\right\}.
    \end{equation}
    \item [2)]  $[\BPhi_j]_{[x, y]} \rightarrow 0$ as $\widetilde{\sigma}^2_e \rightarrow + \infty$, for one of $x$ or $y$ that belongs to $\mathcal{K}_e$. 
\end{itemize}
Here, $[\mathbf{R}_j^{\mathcal{K}_g}]_{[a, b]}$ is the submatrix obtained by partitioning the rows and columns based on $\Bh^{\mathcal{K}_g}_j$, with a size of $k_a \times k_b$. Thus, we have $[\BR_{j}^{\mathcal{K}_g}]_{[a, b]} = [\mathbf{R}_j]_{[k_a, k_b]}$ and $[\BD_{\widetilde{\sigma}}^{\mathcal{K}_g}]_{[a, a]} = \widetilde{\sigma}_{k_a}^2 \BI_{N_{k_a}}$ for each $a, b \in [G]$.  
\par
We prove item 1) first. We only prove the case with $\BR = \BW$ as the proofs for the other terms are similar. For $a, b \in [G]$, we have
\begin{equation}
\begin{split}
\label{Eq_Wj}
    &[\BW_j]_{[k_a, k_b]} =[\BR_j(\BD_{\widetilde{\sigma}} + \BR_j)^{-1}\BD_{\widetilde{\sigma}}]_{[k_a, k_b]} \\
    & = \widetilde{\sigma}_{k_b}^2  [\BR_j(\BD_{\widetilde{\sigma}} + \BR_j)^{-1}]_{[k_a, k_b]}.
\end{split}
\end{equation}
 Without loss of generality, assume $k_a = a$ for $a \in [G]$ and write 
\begin{equation}
    \BR_j = \begin{bmatrix}
        \BR_j^{\mathcal{K}_g} & \BR_{j, 1} \\
        \BR_{j, 2} & \BR_{j, 3}
    \end{bmatrix}, \BD_{\widetilde{\sigma}} = \begin{bmatrix}
         \BD_{\widetilde{\sigma}}^{\mathcal{K}_g} & \mathbf{0} \\
         \mathbf{0} & \BD_{\widetilde{\sigma}, 1}
     \end{bmatrix}.
\end{equation}
According to the block matrix inversion formula \cite{horn2012matrix}, we can get
 \begin{equation}
     (\BD_{\widetilde{\sigma}} + \BR_j)^{-1} = \begin{bmatrix}
         \BP_{1} & \BP_{2} \\
         \BP_{3} & \BP_{4}
     \end{bmatrix},
 \end{equation}
 where terms $\BP_t$, $t = 2, 3, 4$ approach $\mathbf{0}$ as $\widetilde{\sigma}_e^2 \rightarrow + \infty$, and the term
 \begin{equation}
 \begin{split}
      \BP_1 &= (\BR_j^{\mathcal{K}_g} +\BD_{\widetilde{\sigma}}^{\mathcal{K}_g} - \BR_{j, 2}(\BR_{j, 3} + \BD_{\widetilde{\sigma}, 1})^{-1}\BR_{j, 1} )^{-1} \\
      &\rightarrow (\BR_j^{\mathcal{K}_g} +\BD_{\widetilde{\sigma}}^{\mathcal{K}_g})^{-1}, \text{  as  } \widetilde{\sigma}_e^2 \rightarrow + \infty.
 \end{split}
 \end{equation}
Hence, by \eqref{Eq_Wj}, the following holds
\begin{equation}
\begin{split}
    &\lim_{\widetilde{\sigma}_e^2 \rightarrow + \infty}[\BW_j]_{[k_a, k_b]} 
    = \widetilde{\sigma}_{k_b}^2  \left[\begin{bmatrix}
        \BR_j^{\mathcal{K}_g}(\BR_j^{\mathcal{K}_g} +\BD_{\widetilde{\sigma}}^{\mathcal{K}_g})^{-1} & \mathbf{0} \\
        \BR_{j, 2}(\BR_j^{\mathcal{K}_g} +\BD_{\widetilde{\sigma}}^{\mathcal{K}_g})^{-1} & \mathbf{0}
    \end{bmatrix}\right]_{[k_a, k_b]} \\
    &=   [\BR_j^{\mathcal{K}_g}(\BR_j^{\mathcal{K}_g} +\BD_{\widetilde{\sigma}}^{\mathcal{K}_g})^{-1}]_{[a, b]} [\BD_{\widetilde{\sigma}}^{\mathcal{K}_g}]_{[b, b]} = [\BW_j^{\mathcal{K}_g}]_{[a, b]}.
\end{split}
\end{equation}
By summing over the index $j$ in above equation, we get $\lim_{\widetilde{\sigma}_e^2 \rightarrow + \infty}[\BW]_{[k_a, k_b]} = [\BW^{\mathcal{K}_g}]_{[a, b]}$. Therefore, item 1) is proved.
\par
Next, we prove item 2). Without loss of generality, we assume $x \in \mathcal{K}_e$. As a result, 
\begin{equation}
    [\BD_{T, j}]_{[x, x]} = [\BR_j]_{[x, x]} (\widetilde{\sigma}_x^2 \BI_{N_x} + [\BR_j]_{[x, x]})^{-1} \rightarrow \mathbf{0},
\end{equation}
as $\widetilde{\sigma}_e^2 \rightarrow + \infty$.
Hence, we have
\begin{equation}
    [\BPhi_j]_{[x, y]} =  [\BD_{T, j}]_{[x, x]} [(\BD_{\widetilde{\sigma}} + \BR_j)\BD_{T, j}]_{[x, y]} \rightarrow \mathbf{0}.
\end{equation}
Therefore, we complete the proof of Corollary \ref{Coro_Unbalance_CSI}. \qed
}
\vspace{-1mm}
\section{Proof of Lemma \ref{Thm_DE}}
\label{App_Thm_DE}
The proof relies on Gaussian tools \cite{pastur2011eigenvalue, Hachem2008ANewApproach, Xin2023Double-Scattering, Abla2009Ber, zhuang2024twohop}, which consist of the integration by parts formula \cite[Eq. (17)]{Hachem2008ANewApproach} and the Poincaré-Nash  inequality \cite[Eq. (18)] {Hachem2008ANewApproach}. According to the Poincaré–Nash inequality, the variance of functions defined in Table \ref{Table_of_notations} can be proved of order $\BO(1)$ (See \cite[Lemma 9]{Abla2016Noeigen} for the case of $\digamma_k$ and the proof of other functions are similar). As a result, for given $\epsilon > 0$, we have
\setcounter{equation}{89}
\begin{equation}
\begin{split}
    &\sum_{N = 1}^{+ \infty} \mathbb{P}\left(\frac{\abs{\mathcal{X} - \E \mathcal{X}}}{N}\geq \epsilon \right)   \leq \sum_{N = 1}^{+ \infty} \frac{\Var(\mathcal{X})}{\epsilon^2N^2} {<} + \infty,
\end{split}
\end{equation}
where $\mathcal{X}$ can be any function in Table \ref{Table_of_notations}. Thus, according to Borel–Cantelli lemma, we have 
\begin{equation}
\frac{\mathcal{X} - \E 
\mathcal{X}}{N} \xrightarrow[N \xrightarrow{c_1, \ldots, c_K} + \infty]{a.s.} 0.
\end{equation}
Therefore, to show the stochastic equivalence, we only need to evaluate the means of functions in Table \ref{Table_of_notations}.
\subsection{Evaluation of $\E \digamma_k$}
Before delving into the details of the proof, we first define the following quantities 
\begin{align}
    &\alpha_{jk} = \frac{1}{N_k} \Tr [\BOmega_j]_{[k, k]} \BQ_k , ~~ \BD_{\alpha, k} = \diag(\alpha_{jk}; j \in [M] ), \notag\\
    & \widetilde{\BF}_{\alpha, k} = (\BI + \E \BD_{\alpha, k})^{-1}, \notag\\
    &\BTheta_{\alpha, k} = \left(-z_k\BI_{N_k} + \BS_k + \frac{1}{N_k} \sum_{j =1}^M \frac{[\BOmega_j]_{[k, k]}}{1 + \underline{\alpha}_{jk}} \right)^{-1}.
\end{align}
These quantities will serve as intermediate variables in the proof, i.e., we will show $\BQ_k \approx \BTheta_{\alpha, k} \approx \BTheta_k$.
\par
Applying integration by parts formula  to $\E [\BQ_k]_{i, p} [\Bx_s]_a [\BY_l]_{j, s}^*$ and taking $[\Bx_s]_a$ as the variable, we have
\begin{equation}
\label{Eq_DE_Q_1}
    \begin{split}
        & \E [\BQ_k]_{i, p} [\Bx_s]_a [\BY_l]_{j, s}^* =  \E \frac{\partial [\BQ_k]_{i, p}  [\BY_l]_{j, s}^*}{\partial [\Bx_s]_a^*} \\
        &= \E -\frac{1}{\sqrt{N_k}}[\BQ_k \BX_k]_{i, s} [\BA_{sk}^H \BQ_k]_{a, p} [\BY_l]_{j, s}^* \\
        &+ \E \frac{1}{\sqrt{N_l}}[\BQ_k]_{i, p} [\BB_{sl}^H]_{a, j}.
    \end{split}
\end{equation}
Multiplying $\frac{1}{\sqrt{N_k}}[\BA_{sk}]_{p, a}$ at both sides of \eqref{Eq_DE_Q_1} and summing over the subscripts $p$ and $a$, the following holds
\begin{equation}
    \begin{split}
    \label{Eq_QXY}
        &\E [\BQ_k \BX_k]_{i, s} [\BY_l]_{j, s}^* =  \E \frac{[\BQ_k\BA_{sk}\BB_{sl}^H]_{i, j}}{\sqrt{N_kN_l}} \\
        & \hspace{3mm} -\E\alpha_{sk}[\BQ_k\BX_k ]_{i, s} [\BY_l]_{j, s}^* . 
    \end{split}
\end{equation}
By writing the random variable $\alpha_{sk}$ as the sum of its mean and its centered version, i.e., $\alpha_{sk} = \underline{\alpha}_{sk} + \mathring{\alpha}_{sk}$ and solving \eqref{Eq_QXY} w.r.t. $\E [\BQ_k \BX_k]_{i, s} [\BY_l]_{j, s}^*$, we get
\vspace{-1mm}
\begin{equation}
    \begin{split}
    \label{DE_BQHZ}
        &\E [\BQ_k \BX_k]_{i, s} [\BY_l]_{j, s}^* =  \E \frac{[\BQ_k\BA_{sk}\BB_{sl}^H]_{i, j}}{\sqrt{N_kN_l}(1 + \underline{\alpha}_{sk})} \\
        &- \E  \frac{\mathring{\alpha}_{sk}}{1 + \underline{\alpha}_{sk}} [\BQ_k \BX_k ]_{i, s} [\BY_l]_{j, s}^*.
    \end{split}
\end{equation}
\eqref{DE_BQHZ} will also be used for the evaluation of $\E \Phi_{B, kl}$ later. To obtain the approximation of the trace of the resolvent, we set $l = k$ and $\BY_k = \BX_k$, i.e., $\BB_{s} = \BA_{s}$, $\forall s \in [M]$. By summing over the subscript $s$, we can obtain
\begin{equation}
    \begin{split}
        &\E [\BQ_k \BX_k \BX_k^H]_{i, j} =  \E [\BQ_k\sum_{s=1}^M\frac{[\BOmega_{s}]_{[k, k]}}{N_k(1 + \underline{\alpha}_{sk})}]_{i, j} \\
        &- \E [\BQ_k \BX_k \mathring{\BD}_{\alpha, k} \widetilde{\BF}_{\alpha, k} \BX_k^H]_{i, j}.
    \end{split}
\end{equation}
By using the resolvent identity $\BQ_k \BX_k \BX_k^H = z_k \BQ_k -  \BQ_k\BS_k + \BI_{N_k}$ on $\E [\BQ_k \BX_k \BX_k^H]_{i, j}$, multiplying both sides $[\BTheta_{\alpha, k}\BBA]_{j, i}$ , and summing over the subscripts $i$ and $j$,  we can get
\begin{equation}
\label{Eq_AQ_1}
\begin{split}
    &\E \Tr \BBA \BQ_k - \Tr \BBA \BTheta_{\alpha, k} \\
    &=    \E  \Tr \BQ_k \BX_k \mathring{\BD}_{\alpha, k} \widetilde{\BF}_{\alpha, k} \BX_k^H \BTheta_{\alpha, k} \BBA :=   \varepsilon_k(\BBA). 
\end{split}
\end{equation}
We can write the residual term $\varepsilon_k(\BBA)$ as 
\begin{equation}
\begin{split}
    \varepsilon_k(\BBA) &= \sum_{j=1}^M \E  \mathring{\alpha}_{jk} \underbrace{[\widetilde{\BF}_{\alpha, k} \BX_k^H \BTheta_{\alpha, k} \BBA\BQ_k \BX_k]_{j,j}}_{:= \chi_{jk}} \\
    &\overset{(a)}{\leq} \sum_{j=1}^M \Var^{\frac{1}{2}}({\alpha}_{jk})\Var^{\frac{1}{2}}({\chi}_{jk}) \overset{(b)}{=} \BO(\frac{1}{N^{\frac{1}{2}}}), 
\end{split}
\end{equation}
where step $(a)$ follows from Cauchy-Schwartz inequality. Step $(b)$ follows from the fact $\Var(\alpha_{jk}) =  \BO(\frac{1}{N^2})$ and $\Var(\chi_{jk}) = \BO(\frac{1}{N})$, which can be proved by Poincaré-Nash  inequality. Hence, we build the relation between $\BQ_k$ and $\BTheta_{\alpha, k}$.
\par
To establish the relationship between $\BTheta_{\alpha, k}$ and $\BTheta_k$, we take the difference between $\frac{1}{N_k}\Tr [\BOmega_{j}]_{[k, k]} \BTheta_{\alpha, k}$ and $\delta_{jk}$ for each $j \in [M]$. Following the same procedure as in the proof of \cite[Proposition 3]{Abla2016Noeigen}, we can show that
$
    \abs{\Tr \BBA \BTheta_k - \Tr \BBA \BTheta_{\alpha, k}} = \BO(N^{-\frac{1}{2}}),
$
which concludes \eqref{DE_digamma}.
\subsection{Evaluation of $\E\Phi_{B, kl}$}
Multiplying $[\Bb]_s$ and $[\BBA]_{j, i}$ at both sides of \eqref{DE_BQHZ} and summing over the subscripts $s$, $i$, and $j$, we can obtain
\vspace{1mm}
\begin{equation}
    \begin{split}
        &\E \Phi_{B, kl}(\BBA, \Bb) = -\E [\BQ_k \BX_k \diag(\Bb) \mathring{\BD}_{\alpha, k} \widetilde{\BF}_{\alpha, k}\BY_l^H] \\
        &+  \sum_{s}\frac{[\Bb]_s \Tr \BBA (\E \BQ_k) \BA_{sk}\BB_{sl}^H}{\sqrt{N_kN_l}(1 + \underline{\alpha}_{sk})} \\
        &\overset{(a)}{=} \sum_{s}\frac{[\Bb]_s \Tr \BBA \BTheta_k \BA_{sk}\BB_{sl}^H}{\sqrt{N_kN_l}(1 + \delta_{sk})} + \BO(N^{-\frac{1}{2}}),
    \end{split}
\end{equation}
\vspace{1mm}
where step $(a)$ follows from \eqref{DE_digamma} and the variance control.
\vspace{-6mm}
\subsection{Evaluation of $\E \Upsilon_{kl}$}
According to integration by parts formula, we get
\begin{equation}
\begin{split}
&\E [\BQ_k \BBB \BQ_l\BX_l]_{i, s} [\BX_l]_{j, s}^* =   \E \frac{[\BQ_k \BBB \BQ_l[\BOmega_{s}]_{[l, l]}]_{i, j}}{N_l(1 + \underline{\alpha}_{sl})}  \\
&-\E \frac{\mathring{\alpha}_{sl}}{1 + \underline{\alpha}_{sl}} [\BQ_k \BBB \BQ_l \BX_l]_{i, s}  [\BX_l]_{j, s}^* \\
&- \frac{ \E \Upsilon_{kl}( [\BOmega_{s}]_{[l, k]}, \BBB) [\BQ_k \BX_k]_{i, s}  [\BX_l]_{j, s}^*}{\sqrt{N_lN_k}(1 + \underline{\alpha}_{sl})}. 
\end{split}
\end{equation}
Summing over the subscript $s$ and using the resolvent identity $\BQ_l \BX_l \BX_l^H = z_l \BQ_l -  \BQ_l\BS_l + \BI_{N_l}$,  the following holds
\vspace{2mm}
\begin{equation}
\begin{split}
&\E [\BQ_k \BBB \BQ_l \BTheta_{\alpha, l}^{-1}]_{i, j}  =  \E[\BQ_k\BBB]_{i, j} \\
&+ \E [\BQ_k \BX_k\BD_{\Upsilon, kl}(\BBB)\widetilde{\BF}_{\alpha, l}\BX_l^H]_{i, j} + [\boldsymbol{\varepsilon}_{kl}]_{i, j}, \vspace{2mm}
\end{split}
\end{equation}
where $\BD_{\Upsilon, kl}(\BBB) = \frac{1}{\sqrt{N_kN_l}}\diag({\Upsilon_{kl}( [\BOmega_{s}]_{[l, k]}, \BBB)}; s \in [M])$ and  $\boldsymbol{\varepsilon}_{kl}$ is the residual matrix. It can be proved by Poincaré–Nash inequality that $ \Tr \boldsymbol{\varepsilon}\BBC = \BO(N^{-\frac{1}{2}})$ for any $\BBC$ with bounded spectral norm. Thus, we have
\vspace{2mm}
\begin{equation}
\begin{split}
\label{Eq_DE_Upsilon_AB_O}
&\E \Upsilon_{kl}( \BBA, \BBB)  =  \Tr \BBA (\E\BQ_k)\BBB\BTheta_{\alpha, l} \\
&+ \E\Tr  \BQ_k \BX_k {\BD}_{\Upsilon, kl}(\BBB)\widetilde{\BF}_{\alpha, l}\BX_l^H \BTheta_{\alpha, l} \BBA + \BO(N^{-\frac{1}{2}}). 
\end{split}
\end{equation}
By replacing $\BTheta_{\alpha, l}$ and $\widetilde{\BF}_{\alpha, l}$ with $\BTheta_{l}$ and $\widetilde{\BF}_{l}$, respectively, setting $\BBA = [\BOmega_{u}]_{[l, k]}$ and $\BBB = [\BOmega_{v}]_{[k, l]}$, and using the approximation rules of $\Phi_{B, kl}$ in \eqref{DE_Phi}, we have
\begin{equation}
\label{Eq_DE_Upsilon_AB_1}
    \begin{split}
        &[{\BUpsilon}_{\Omega, kl}]_{u, v} = N_kN_l [\BGamma_{kl}]_{u, v} \\
        &+ \sum_{j=1}^M [{\BUpsilon}_{\Omega, kl}]_{j, v} [\widetilde{\BF}_k \widetilde{\BF}_l]_{j, j} [\BGamma_{kl}]_{u, j} + \BO(N^{-\frac{1}{2}}),
    \end{split}
\end{equation}
where $[{\BUpsilon}_{\Omega, kl}]_{u, v} = \E \Tr [\BOmega_{u}]_{[l, k]} \BQ_k [\BOmega_{v}]_{[k, l]} \BQ_l$. Solving \eqref{Eq_DE_Upsilon_AB_1} w.r.t. 
${\BUpsilon}_{\Omega, kl}$, we get
\begin{equation}
    \begin{split}
        {\BUpsilon}_{\Omega, kl} &= N_kN_l \BXi_{kl}\BGamma_{kl} + \BO(N^{-\frac{1}{2}}) \mathbf{1}_{M} \mathbf{1}_M^T.
    \end{split}
\end{equation}
By following the similar method as in \cite{Abla2016Noeigen, Hachem2008CLTVP}, we can show that $\BI_{M} - \BGamma_{kl}\widetilde{\BF}_{k} \widetilde{\BF}_l $ is invertible and $ \max_{i \in [M]} \sum_{j=1}^M \abs{[\BXi_{kl}]_{i,j}} = \BO(1)$. Due to the space limitation, we omit the details. Similar to the evaluation of ${\BUpsilon}_{\Omega, kl}$, we can derive the approximation rules for $\E \Upsilon_{kl}( \BBA, [\BOmega_{v}]_{[k, l]} )$ and $\E \Upsilon_{kl}( [\BOmega_{u}]_{[l, k]}, \BBB )$. Plugging these results in  \eqref{Eq_DE_Upsilon_AB_O} yields
\begin{equation}
\begin{split}
      &\E \Tr \BBA \BQ_k \BBB \BQ_l = \Tr \BBA \BTheta_k \BBB \BTheta_l \\
      &+  \widetilde{\Blambda}_{kl}( \BBA) \widetilde{\BF}_{k} \widetilde{\BF}_{l} {\BXi}_{kl}{\Blambda}_{kl}(\BBB) + \BO(N^{-\frac{1}{2}}),
\end{split}
\end{equation}
which proves \eqref{DE_Upsilon}.
\vspace{-2mm}
\subsection{Evaluation of $\E \Pi_{B, kl}$}
By applying integration by parts formula, we can obtain
\begin{equation}
\begin{split}
    &\E  [\BY_l^H \BQ_l \BBA \BQ_k]_{i, p} [\BY_k]_{p, i} =  \E \Big\{ \frac{ [\BB_{ik}\BB_{il}^H \BQ_l \BBA \BQ_k]_{p, p}}{\sqrt{N_kN_l}} \\
    &- \frac{1}{\sqrt{N_kN_l}}[\BY_l^H\BQ_l \BX_l]_{i, i}[\BB_{ik}\BA_{il}^H \BQ_l \BBA \BQ_k]_{p, p} \\
    &- \frac{1}{N_k}[\BY_l^H \BQ_l \BBA \BQ_k \BX_k ]_{i, i} [\BB_{ik}\BA_{ik}^H \BQ_k]_{p, p} \Big\}. 
\end{split}
\end{equation}
Summing over the subscript $p$, and using the variance control and the approximation rules for $\Upsilon_{kl}$, we get
\vspace{2mm}
\begin{equation}
\label{Eq_DE_ZQAQZ_1}
\begin{split}
    &\E  [\BY_l^H \BQ_l \BBA \BQ_k\BY_k]_{i, i} 
    = \frac{\overline{\Upsilon}_{kl}(\BBA, \BB_{ik}\BB_{il}^H) }{\sqrt{N_kN_l}} \\
    &- \frac{\overline{\Upsilon}_{kl}(\BBA, \BB_{ik}\BA_{il}^H)}{\sqrt{N_kN_l}} \E[\BY_l^H\BQ_l \BX_l]_{i, i} \\
    &- \E [\BY_l^H \BQ_l \BBA \BQ_k \BX_k ]_{i, i} \frac{ \Tr \BB_{ik}\BA_{ik}^H \BTheta_k}{N_k}+ \BO(N^{-\frac{3}{2}}).
\end{split}
\end{equation}
\vspace{2mm}
\setcounter{equation}{108}
\begin{figure*}[t]
\begin{equation}
\begin{split}
\label{Eq_ZQAQZ}
    &\E \Pi_{B, kl} (\BBA) 
    =  \sum_{i=1}^M  \Big\{ \Big[ \widetilde{\Blambda}_{B, B, kl}(\BBA) +  \widetilde{\Blambda}_{kl}( \BBA) \widetilde{\BF}_{k} \widetilde{\BF}_{l} {\BXi}_{kl}{\BLambda}_{B, B, kl}  - [\BD_{ A, B, l} \widetilde{\BF}_{l}]_{i, i} \Big(  \widetilde{\Blambda}_{B, A, kl}( \BBA) +  \widetilde{\Blambda}_{kl}( \BBA) \widetilde{\BF}_{k} \widetilde{\BF}_{l} {\BXi}_{kl}{\BLambda}_{B, A, kl} \Big) \\
    &- [\BD_{B, A, k} \widetilde{\BF}_{k}]_{i, i} \Big(\widetilde{\Blambda}_{A, B, kl}(\BBA) +  \widetilde{\Blambda}_{kl}(\BBA) \widetilde{\BF}_{k} \widetilde{\BF}_{l} {\BXi}_{kl}{\BLambda}_{A, B, kl}\Big) \Big] \Be_i + [\BD_{B, A, k} \BD_{A, B, l} \widetilde{\BF}_{k}\widetilde{\BF}_{l}]_{i, i} [\widetilde{\Blambda}_{kl}( \BBA) \widetilde{\BXi}_{kl}]_{i}\Big\}  + \BO(N^{-\frac{1}{2}})\\
    &\overset{(a)}{=}  \Big( \widetilde{\Blambda}_{B, B, kl}(  \BBA) +  \widetilde{\Blambda}_{kl}( \BBA) \widetilde{\BF}_{k} \widetilde{\BF}_{l} {\BXi}_{kl}{\BLambda}_{B, B, kl} - \widetilde{\Blambda}_{B, A, kl}(  \BBA)\BD_{A, B, l} \widetilde{\BF}_{l} - \widetilde{\Blambda}_{kl}( \BBA) \widetilde{\BF}_{k} \widetilde{\BF}_{l} {\BXi}_{kl}{\BLambda}_{B, A, kl}\BD_{A, B, l} \widetilde{\BF}_{l} \\
    &- \widetilde{\Blambda}_{A, B, kl}( \BBA)\BD_{B, A, k} \widetilde{\BF}_{k}  -  \widetilde{\Blambda}_{kl}( \BBA) \widetilde{\BF}_{k} \widetilde{\BF}_{l} {\BXi}_{kl}{\BLambda}_{A, B, kl}\BD_{ B, A, k} \widetilde{\BF}_{k}  +   \widetilde{\Blambda}_{kl}( \BBA) \widetilde{\BF}_{k} \widetilde{\BF}_{l} {\BXi}_{kl}\BD_{B, A, k} \BD_{A, B, l} \Big)\mathbf{1}_M  + \BO(N^{-\frac{1}{2}}).
\end{split}
\end{equation}
\hrulefill
\vspace{-3mm}
\end{figure*}
Similarly, we can obtain
\setcounter{equation}{107}
\begin{equation}
\begin{split}
\label{Eq_YQAQX}
    & \E  [\BY_l^H \BQ_l \BBA \BQ_k\BX_k]_{i, i} =  \frac{ \overline{\Upsilon}_{kl}(\BBA, \BA_{ik}\BB_{il}^H)}{\sqrt{N_kN_l}(1 + \delta_{ik})} \\
    &- \frac{\overline{\Upsilon}_{kl}( \BBA, [\BOmega_{i}]_{[k, l]})}{\sqrt{N_kN_l}(1 + \delta_{ik})} \E[\BY_l^H\BQ_l \BX_l]_{i, i}+ \BO(N^{-\frac{3}{2}}).
\end{split}
\end{equation}
Then, plugging \eqref{Eq_YQAQX} into \eqref{Eq_DE_ZQAQZ_1}, summing over the subscript $i$ and using the approximation rules for $\Phi_{B, kl}$ in \eqref{DE_Phi}, we get \eqref{Eq_ZQAQZ} at the top of the next page, where $\widetilde{\BXi}_{kl} = \BI_M + \widetilde{\BF}_k\widetilde{\BF}_l\BXi_{kl}\BGamma_{kl}$, and step $(a)$ in \eqref{Eq_ZQAQZ} follows by the identity $\widetilde{\BF}_k\widetilde{\BF}_l \BXi_{kl} = \widetilde{\BXi}_{kl} \widetilde{\BF}_k\widetilde{\BF}_l  $. Therefore, we complete the proof of Lemma \ref{Thm_DE}. \qed
\section{Proof of Corollary \ref{Coro_best_rho}}
\label{App_Coro_best_rho}
According to Corollary \ref{IID_Case}, we know that $\delta_k$ is the positive solution of the following equation
\setcounter{equation}{109}
\begin{equation}
    \delta_k(\rho_k) = \frac{1}{\rho_k(\widetilde{\sigma}^2+1) + \frac{M+1}{N_k} \widetilde{\sigma}^2 + \frac{M}{N_k} \frac{1}{1 + \delta_k(\rho_k)}}.
\end{equation}
By taking the derivative w.r.t. $\rho_k$, we have
\begin{equation}
\label{Eq_D_delta_k}
    \frac{\partial \delta_k}{\partial \rho_k} = -\frac{(\widetilde{\sigma}^2 + 1) \delta_k^2}{V_k}, 
\end{equation}
where $V_k = 1 - \frac{M}{N_k} \frac{\delta_k^2}{(1 + \delta_k)^2}$ and it can be proved that $V_k > 0$. Plugging this result into \eqref{gamma_opt} yields
\begin{equation}
    \overline{\gamma}^{\mathrm{lfoc}}_K = \sum_{k=1}^K\frac{\delta_k^2}{\delta_k + (\rho_k - \frac{\sigma^2}{N_k}) \frac{\partial \delta_k}{\partial \rho_k}} := \sum_{k=1}^K \overline{\gamma}^{\mathrm{lfoc}}_{K, k}.
\end{equation}
By taking the derivative, we obtain
\begin{equation}
    \frac{\partial \overline{\gamma}^{\mathrm{lfoc}}_K}{\partial \rho_k} = \frac{\partial \overline{\gamma}^{\mathrm{lfoc}}_{K, k}}{\partial \rho_k} =  \frac{\left[2(\frac{\partial \delta_k}{\partial \rho_k} )^2 - \delta_k \frac{\partial^2 \delta_k}{\partial \rho_k^2}\right](\rho_k - \frac{\sigma^2}{N_k})\delta_k}{\left[\delta_k + (\rho_k - \frac{\sigma^2}{N_k}) \frac{\partial \delta_k}{\partial \rho_k}\right]^2}.
\end{equation}
By taking the second-order derivative of $\delta_k$, we have
\begin{equation}
    2\left(\frac{\partial \delta_k}{\partial \rho_k} \right)^2 - \delta_k \frac{\partial^2 \delta_k}{\partial \rho_k^2} = -\frac{2M(\widetilde{\sigma}^2+1)^2 \delta_k^6}{N_kV_k^3(1 + \delta_k)^3} < 0.
\end{equation}
Hence, the optimal solution is given by $\rho_k^* = \frac{\sigma^2}{N_k} = \arg \max_{\rho_k > 0} \overline{\gamma}^{\mathrm{lfoc}}_{K, k}$. 
Since $\rho_k$ is independent with $\gamma^{\mathrm{lfoc}}_{K, l}$  for $l \neq k$,
the optimal $\Brho_K$ is given by $ \Brho_K^* = [\frac{\sigma^2}{N_1}, \ldots, \frac{\sigma^2}{N_K}]^T$, which concludes \eqref{Eq_best_rhoK}. \qed
\section{Proof of Corollary \ref{Coro_best_c}}
\label{App_Coro_best_c} 
Let $\rho_k = \frac{a}{c_k}$, $\forall k$ and define $A = a(\widetilde{\sigma}^2 + 1) + \frac{ M + 1}{M}\widetilde{\sigma}^2$ and $f(x) = Ax + \frac{x}{1 + x}$. Then we consider the following equation
 \begin{equation}
 \label{Eq_f}
     f(x) = c,
 \end{equation}
 with $c \geq 0$. It can be proved that \eqref{Eq_f} has a unique positive solution $x = \delta(c)$ and we can observe that $\delta_k = \delta(c_k)$. Furthermore, we define 
 \begin{equation}
 \label{Eq_wgamma_c}
    \widetilde{\gamma}(c) = \frac{\delta(c) }{1 + B \frac{ \delta(c)}{c -  \frac{\delta^2(c)}{(1 + \delta(c))^2}}},
 \end{equation}
 where $B = (\frac{\sigma^2}{M} - a)(\widetilde{\sigma}^2 + 1)$. Then we can obtain $\overline{\gamma}^{\mathrm{lfoc}}_K(\Brho_K, \Bc_K) = \sum_{k=1}^K \widetilde{\gamma}(c_k)$.
Taking the derivative of $f(\delta(c)) = c$ w.r.t. $c$ yields
\begin{align}
\label{Eq_partial_delta}
    \frac{\partial \delta}{\partial c} &= \frac{1}{A + \frac{1}{(1 + \delta)^2}}, ~~  \frac{\partial^2 \delta}{\partial c^2} =  \frac{2(\frac{\partial \delta}{\partial c})^3}{(1 + \delta)^3}, \notag \\
    \frac{\partial^3 \delta}{\partial c^3} &= \frac{6 (\frac{\partial \delta}{\partial c})^4}{(1 + \delta)^4}\left( \frac{\partial \delta}{\partial c}\frac{2}{(1 + \delta)^2} - 1\right),
\end{align}
where $\delta = \delta(c)$. Next we will prove \eqref{Eq_gamma_c_M}. By taking the sum $\sum_{k=1}^K c_k = \sum_{k=1}^K f(\delta_k)$, we have
\begin{equation}
     K\overline{c} = A\sum_{k=1}^K {\delta_k} + \sum_{k=1}^K \frac{\delta_k}{1 + \delta_k} \overset{(a)}{\geq}  f(\sum_{k=1}^K \delta_k),
\end{equation}
where step $(a)$ follows from the inequality $\frac{x}{1+x} + \frac{y}{1 + y} \geq \frac{x+y}{1 + x + y}$ for $x, y \geq  0$. Since $f$ is monotonically increasing, we have $\sum_{k=1}^K \delta_k \leq \delta(K\overline{c})$. Through basic transformation, we can obtain
\begin{equation}
\begin{split}
\label{Eq_Pdelta_Pc}
    \frac{\delta}{c - \frac{\delta^2}{(1+\delta^2)}}  = \frac{1}{A + \frac{1}{(1 + \delta)^2}} = \frac{\partial \delta}{\partial c}.
\end{split}
\end{equation}
By \eqref{Eq_wgamma_c} and \eqref{Eq_Pdelta_Pc}, the following holds
\begin{equation}
    \sum_{k=1}^K \widetilde{\gamma}(c_k) = \sum_{k=1}^K \frac{\delta(c_k)}{1 + B \frac{\partial\delta(c_k) }{\partial c}} \overset{(a)}{\leq }  \frac{\sum_{k=1}^K \delta(c_k)}{1 + B \frac{\partial\delta(K\overline{c}) } {\partial c}} \leq \widetilde{\gamma}(K\overline{c}),
\end{equation}
where step $(a)$ follows from $B \leq 0$ for $a \geq \frac{\sigma^2}{M}$ and $\frac{\partial^2 \delta}{\partial c^2} \geq 0$. Therefore, $\eqref{Eq_gamma_c_M}$ holds.
\par
Next, we assume that $ \frac{\sigma^2}{M} \leq a \leq \frac{2 \sigma^2}{M} + \frac{(M+1)\widetilde{\sigma}^2}{M(\widetilde{\sigma}^2 + 1)}$. We will first prove $\widetilde{\gamma}(c)$ is convex. For that purpose, we take the second-order derivative of $\widetilde{\gamma}$ w.r.t. $c$ as
\begin{equation}
\begin{split}
    \frac{\partial^2 \widetilde{\gamma}}{\partial c^2} &= \frac{1}{(1 + B \frac{\partial \delta}{\partial c})^2} \Big\{-(B^2\delta \frac{\partial \delta}{\partial c} + B\delta)\frac{\partial^3 \delta}{\partial c^3} \\
    &+ 2B^2\delta(\frac{\partial^2 \delta}{\partial c^2})^2 - (B^2(\frac{\partial \delta}{\partial c})^2 - 1)\frac{\partial^2 \delta}{\partial c^2}\Big\}. \label{Eq_par_wgamma_par_c}
\end{split}
\end{equation}
Thus, to prove $\frac{\partial^2 \widetilde{\gamma}}{\partial c^2} \geq 0$, it is sufficient to show that
\begin{equation}
\label{Eq_pq_1}
    (-2pq^2  + 3(1 - 2p) q - 1) \delta \leq 1 - q^2,
\end{equation}
where $p = A\frac{\partial \delta}{\partial c}$ and $q = B\frac{\partial \delta}{\partial c}$. For  $\frac{\sigma^2}{M}  \leq a \leq \frac{2\sigma^2}{M} + \frac{(M+1) \widetilde{\sigma}^2}{M}$, it holds true that $-\frac{1}{2} \leq \frac{B}{A} = \frac{q}{p} \leq 0$. From \eqref{Eq_partial_delta}, it can be observed that $0 \leq p < 1$. Hence, we can obtain
\begin{equation}
    \underset{p \in [0, 1), q \in [-\frac{p}{2}, 0] }{\sup} -2pq^2  + 3(1 - 2p) q - 1 = 0,
\end{equation}
to conclude \eqref{Eq_pq_1}. According to Jensen's inequality, we have
\begin{equation}
 \sum_{k=1}^K \widetilde{\gamma}(c_k)  \geq K \widetilde{\gamma}(\overline{c}) = \overline{\gamma}^{\mathrm{lfoc}}_K(\Brho_K, {\Bc}_{K, \mathrm{m}}),
\end{equation}
which completes the proof of Corollary \ref{Coro_best_c}. \qed
\section{Proof of Corollary \ref{Coro_impact_K}}
\label{App_Coro_impact_K}
In this proof, we will reuse the notations in Appendix \ref{App_Coro_best_c}. Moreover, we define $c_{\mathrm{sum}} = \frac{N}{M}$. By \eqref{Eq_Pdelta_Pc}, the SINR is given by 
\begin{equation}
\overline{\gamma}^{\mathrm{lfoc}}_K(\Brho_K, \Bc_{K, \mathrm{m}}) = \frac{K\overline{\delta}_K}{1 + \frac{B}{A + \frac{1}{(1 + \overline{\delta}_K)^2}}},
\end{equation}
where $\overline{\delta}_K$ is the unique positive solution of the following quadratic equation
\begin{equation}
\label{Eq_rK}
    f(\overline{\delta}_K) = \frac{c_{\mathrm{sum}}}{K}.
\end{equation}
 Since $f(x)$ is monotonically increasing w.r.t. $x$, $\overline{\delta}_K$ monotonically decreases w.r.t. $K$. Additionally, \eqref{Eq_rK} implies 
\begin{equation}
    K\overline{\delta}_K = \frac{c_{\mathrm{sum}}}{A + \frac{1}{1 + \overline{\delta}_K}},
\end{equation}
which further indicates that $K\overline{\delta}_K$ is also monotonically decreasing. Since $a \geq \frac{\sigma^2}{M}$, we have $B \leq 0$ and 
\begin{equation}
\begin{split}
&\overline{\gamma}^{\mathrm{lfoc}}_K(\Brho_K, \Bc_{K, \mathrm{m}})  \geq \frac{(K+1)\overline{\delta}_{K+1}}{1 + \frac{B}{A + \frac{1}{(1 + \overline{\delta}_K)^2}}} \\
&\geq \frac{(K+1)\overline{\delta}_{K+1}}{1 + \frac{B}{A + \frac{1}{(1 + \overline{\delta}_{K+1})^2}}} = \overline{\gamma}^{\mathrm{lfoc}}_{K+1}(\Brho_{K+1}, \Bc_{K+1, \mathrm{m}}),
\end{split}
\end{equation}
which conclude \eqref{Eq_Mono_SINR_K}. \qed

\bibliographystyle{IEEEtran}
\bibliography{reference}

\end{document}